\documentclass{article}

\usepackage[preprint, nonatbib]{neurips_2026}

\usepackage[utf8]{inputenc} 
\usepackage[T1]{fontenc}    
\usepackage{hyperref}       
\usepackage{url}            
\usepackage{booktabs}       
\usepackage{amsfonts}       
\usepackage{nicefrac}       
\usepackage{microtype}      
\usepackage{xcolor}         

\usepackage{svg}
\usepackage{enumitem,kantlipsum}
\usepackage{microtype}
\usepackage{graphicx}
\usepackage{subfigure}
\usepackage{booktabs}
\usepackage{hyperref}

\usepackage{amssymb}
\usepackage{mathtools}
\usepackage{amsthm}
\usepackage[capitalize,noabbrev]{cleveref}
\usepackage{subcaption}
\usepackage{amsmath}
\usepackage{algpseudocode}
\usepackage[textsize=tiny]{todonotes}
\usepackage{algorithm}
\usepackage{algpseudocode}

\theoremstyle{plain}

\theoremstyle{definition}

\theoremstyle{remark}

\title{Optimal photostimulation selection for iterative activity maps}

\author{%
  \textbf{Jacob J.~Morra}$^{\dagger}$\quad
  \textbf{Kaitlyn E.~Fouke}\quad
  \textbf{Owen Traubert}\quad
  \textbf{Eva A.~Naumann}\\[0.4em]
  Duke University, Durham NC 27705, USA\\
  $^{\dagger}$\texttt{jacob.morra@duke.edu}
}

\begin{document}

\maketitle

\begin{abstract}
All-optical two-photon holographic optogenetics enables causal circuit mapping by stimulating defined neurons or ensembles while imaging population activity. Yet exhaustive connectivity mapping remains experimentally prohibitive because of combinatorial complexity, tissue heating, photodamage, and experimental time. We present OPhELIA (Optimal Photostimulation sElection for Iterative Activity maps), a Bayesian framework for selecting informative perturbations under limited trial budgets. OPhELIA combines Beta-Bernoulli connectivity inference with an ambiguity-based acquisition heuristic and learned priors derived from pre-stimulation neural activity, augmenting active learning and compressed sensing. In standalone simulations and \textit{in vivo} larval zebrafish visuomotor experiments, OPhELIA with active learning improves trial-efficient approximation of exhaustive functional connectomes. In combinatorial \textit{in vivo} experiments, OPhELIA with compressed sensing most closely recovers an exhaustive connectome using only 5\% of trials. These results establish OPhELIA as a sample-efficient framework for causal connectomics.
\end{abstract}

\begin{figure}[!h]
    \centering
    \makebox[\textwidth][c]{%
        \includegraphics[width=1.25\textwidth]{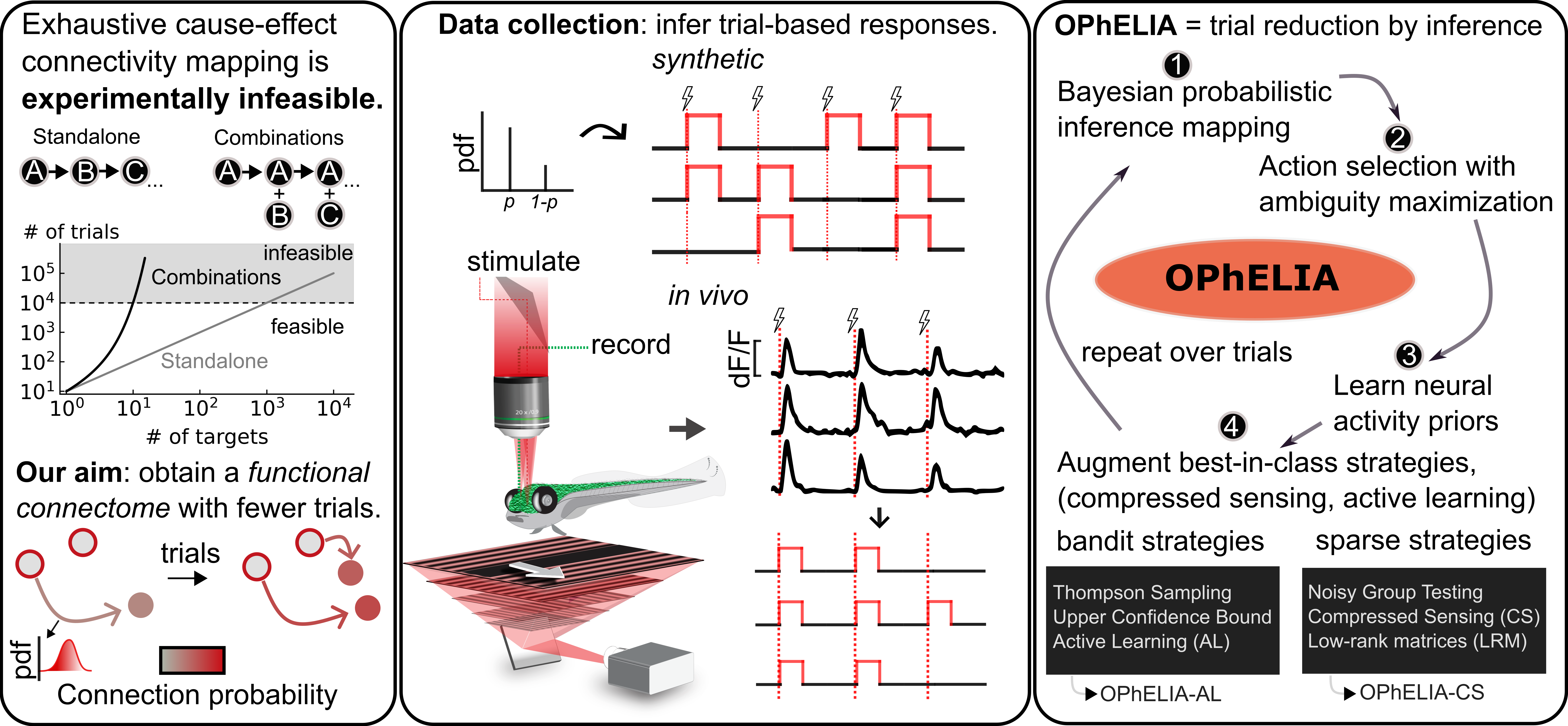}
    }
    \label{fig:ophelia_overview}
\end{figure}

\section{Introduction}
Uncovering how neural activity propagates through circuits at cellular resolution is a fundamental goal in systems neuroscience. All-optical physiology now makes this goal experimentally accessible: two-photon holographic optogenetic photostimulation can activate specified target neurons or ensembles while simultaneously recording large neural populations \cite{Emiliani2015-uk, Yang2018-ey, Marshel2019-yp, Dal-Maschio2017-dp, Chettih2019-vb, Draelos2025-we, Oldenburg2023-ki}, enabling causal circuit discovery of local circuits and functionally-defined sub-networks \cite{Russell2024-ad, Rowland2023-zx}. However, scaling up to systematic causal connectivity mapping remains an open challenge, because the action space grows combinatorially, and because experiments are constrained by photodamage and tissue heating \cite{Picot2018-cj, Forli2021-hb}, photobleaching \cite{Papaioannou2022-rp}, and personnel time.

Existing approaches impose structure on the connectivity inference problem. Compressed sensing recovers sparse connectivity from undersampled measurements \cite{Navarro2023-wa, Chen2025-pu, Triplett2026-ut, Mishchenko2012-uv}; low-rank recovery assumes that connectivity matrices lie in a low-dimensional subspace and fills unobserved entries through matrix completion \cite{Bull2024-vh}; and noisy group testing uses pooled perturbations to recover sparse binary interactions from aggregate responses \cite{Draelos2020-sd, Draelos2020-zj}. These methods are powerful when their assumptions match the underlying circuitry. However, neural connectivity may also be dense, distributed, state-dependent, co-tuned, redundant, or correlated \cite{Svara2022-ik, Stringer2019-xy, Wolf2017-nh, Vishwanathan2024-uv, Petkova2025-ob, Motta2019-mq}. In these settings, the central question is not only how to reconstruct a functional connectome from fewer measurements, but also how to select the next perturbation while the map remains uncertain.

We address this question with \textbf{OPhELIA} (\textbf{O}ptimal \textbf{Ph}otostimulation s\textbf{E}lection for \textbf{I}terative \textbf{A}ctivity maps), an adaptive framework for trial-constrained Bayesian causal connectivity mapping, where a \textbf{trial} is defined as one stimulation event delivered to a specified target neuron or ensemble. OPhELIA treats each stimulated target-responder pair as a probabilistic causal edge and updates its belief over that edge using a Beta-Bernoulli model. This formulation is useful experimentally because it produces interpretable connection probabilities, tracks uncertainty across trials, and can incorporate biological or experimental prior information. Instead of treating all perturbations equally, OPhELIA selects targets whose downstream response probabilities remain ambiguous and thus most informative for reducing disagreement with a trial-exhaustive reference map.

OPhELIA is designed for two experimentally common regimes. In the first, the experimenter has predefined stimulation targets or ensembles, such as genetically or functionally identified cell groups \cite{Naumann2016-mx, Marshel2019-yp, Fouke2025-xy}. We formulate this as a standalone $\mathcal{N}_s$-armed bandit, in which each arm corresponds to a stimulation target. In the second, the experimenter considers combinations of candidate neurons to comprehensively map a circuit. We frame this as a combinatorial selection problem, allowing OPhELIA to augment best-in-class strategies (active learning, compressed sensing) over large ensemble action spaces.

With OPhELIA we present three contributions. First, we introduce a Bayesian connectivity-inference framework that converts repeated photostimulation trials into probabilistic causal maps. Second, we derive an ambiguity-based acquisition heuristic that prioritizes target-responder weights with high posterior uncertainty and response probabilities near maximal Bernoulli entropy. Third, we show how pre-experiment neural activity can be leveraged as a learned biological prior over candidate stimulation targets. Across simulations and \textit{in vivo} larval zebrafish experiments, OPhELIA improves standalone target selection relative to non-adaptive strategies, and in the combinatorial setting, approximates a trial-exhaustive functional connectome with substantially fewer trials. In particular, OPhELIA combined with compressed sensing most closely recovers an exhaustive map using only 5\% of trials, suggesting a practical route toward large-scale causal connectomics.

\section{Related Work}
Two-photon holographic optogenetics now allows experimenter-specified neurons and ensembles to be perturbed while recording large neural populations, enabling causal circuit mapping at cellular resolution \cite{Emiliani2015-uk,Yang2018-ey,Marshel2019-yp,Adesnik2021-review,Oldenburg2023-ki,Draelos2025-we}. This flexibility, however, introduces a combinatorial design problem: exhaustive causal mapping scales poorly with the number of candidate targets and responders. Prior work has therefore used algorithmic sampling strategies to reduce the number of required perturbations, including compressed sensing, noisy group testing, low-rank matrix recovery, and active learning \cite{Candes2006-jn,Navarro2023-wa,Chen2025-pu,Triplett2025-rapid,Draelos2020-sd,Bull2024-vh, Wagenmaker2024-active}.

\textit{Compressed sensing} assumes that connectivity is sparse and recoverable from undersampled measurements \cite{Candes2006-jn,Mishchenko2012-uv,Hu2012-ct,Lee2019-nj,Navarro2023-wa,Chen2025-pu,Triplett2025-rapid}. In all-optical circuit mapping, this is commonly implemented by stimulating random or designed ensembles and reconstructing connectivity through sparse optimization. For example, \cite{Navarro2023-wa} use patterned optogenetic perturbations to infer connectivity from ensemble stimulation, while \cite{Chen2025-pu} and \cite{Triplett2025-rapid} extend sparse reconstruction toward synaptic or monosynaptic mapping. OPhELIA addresses a complementary decision problem: which target or ensemble should be selected next to efficiently approximate a trial-exhaustive functional connectome.

\textit{Noisy group testing} (NGT) uses pooled perturbations to recover sparse binary connectivity under noise. \cite{Draelos2020-sd,Draelos2020-zj} formulate photostimulation-based connectivity estimation as NGT and derive a scalable variational Bayesian approximation to binary-edge posteriors with maximum-entropy regularization. \textit{Low-rank recovery} assumes that population responses admit a compact latent representation, while \textit{active learning} selects perturbations expected to improve estimation of these models \cite{Bull2024-vh,Wagenmaker2024-active}. OPhELIA similarly uses uncertainty-guided selection but estimates graded response probabilities using physiological priors across standalone and combinatorial actions.

Others have sought to optimize holographic photostimulation delivery and parameters. For example, \cite{Triplett2023-pr,Triplett2022-ba,Weiler2024-pk,Triplett2026-ut} focus on laser power and off-target activation. These approaches address the upstream problem of ensuring that the intended target is stimulated accurately. OPhELIA is complementary: it assumes that stimulation has been sufficiently calibrated and focuses on the downstream experimental design problem of which target or ensemble should be selected next. 

Finally, closed-loop all-optical platforms show that neural activity can be analyzed and perturbed in real time, creating the technical basis for adaptive causal mapping \cite{Draelos2021-sb}. However, such platforms do not by themselves define an experimental policy for selecting perturbations. OPhELIA contributes this decision rule by using posterior uncertainty and experimental priors to select informative photostimulation targets.

\vspace{-.15cm}

\section{Methods}
\vspace{-.15cm}
\subsection{Bayesian connectivity inference}\label{ssec:bayfconn}
To iteratively map successive photostimulation trial-based outcomes, we frame causal connectivity as Bayesian inference. Thus, for each stimulated target $i$ to responder $j$, we infer a probability density function, which has an added benefit compared to frequentist probabilistic mappings in that it can natively incorporate rich priors, or be augmented with any number of experimental covariates. 

For a population of $\mathcal{N}_{\rm{s}}$ putative target neurons or ensembles and $\mathcal{N}_{\rm{r}}$ responder neurons, we wish to approximate their causal connectivity ${\theta}_{ij}$ for all $i \in [1,\mathcal{N}_{\rm{s}}]$, $j \in [1,\mathcal{N}_{\rm{r}}]$. We assume that each connectivity probability is Beta distributed, that is $\theta_{ij} \sim \mathrm{Beta}(\alpha, \beta)$. We incorporate updates to this belief using trial data $r_{ij}^{(t)} \sim \rm{Bernoulli}(\theta_{ij})$ for trials $t \in [1,T]$ and $T$ total trials. Using Bayes' Rule to compute the posterior, $\theta_{ij} \text{  } | \text{ } r_{ij}^{(1)} \hdots r_{ij}^{(t)}$, for $\{r_{ij}^{(1)} \hdots r_{ij}^{(t)}\} = \mathcal{D}_{ij}^{(t)}$ we have:  

\begin{equation}
\theta_{ij} \mid \mathcal{D}_{ij}^{(t)} \sim \mathrm{Beta}\left(\alpha_{ij}^{(t)}, \beta_{ij}^{(t)}\right), \text{ where } \alpha_{ij}^{(t)} = \alpha_0 + \sum_{s=1}^{t} r_{ij}^{(s)}, \text{ }
\beta_{ij}^{(t)} = \beta_0 + \sum_{s=1}^{t} (1 - r_{ij}^{(s)}),
\end{equation}
where $r_{ij}^{(t)}=1$ is based on an empirically-derived threshold (see Sec. \ref{sssec:exp_criteria}). The expectation of the posterior is therefore the mean of the distribution:
\begin{equation}
\mathbb{E}[\theta_{ij} \text{ }| \text{ } \mathcal{D}_{ij}^{(t)}]=\hat{\theta}_{ij}^{(t)} = \frac{\alpha_{ij}^{(t)}}{\alpha_{ij}^{(t)} + \beta_{ij}^{(t)}}.
\end{equation}

From each estimate, we define a \textit{causal connectome} as a directed graph $\mathcal{G}$ of each target $i$ to responder $j$, where we assert that causal connectivity is approached after $\mathcal{T}$ exhaustive trials. In matrix notation: 

\begin{equation}
\hat{\Theta}^{(t)} = [\hat{\theta}_{ij}^{(t)}]^{\mathcal{N}_s, \mathcal{N}_r}_{i=1,j=1} \in[0,1]^{\mathcal{N}_s\times\mathcal{N}_r}, \text{ where }\Theta^{(t)}   
\xrightarrow[]{\mathcal{T}\gg t}\Theta.
\end{equation}

This assertion would be unnecessary in simulations only, but is particularly important for experimental validation, since, for example, there is no ground truth functional connectome for the larval zebrafish.

\subsection{Optimized connectivity mapping}\label{ssec:optim}
We propose the following \textbf{problem}: Let true connectome $\Theta$ be approximated as
\begin{equation}
\Theta \approx \Theta^{(\mathcal{T})}
=
\left\{
\mathrm{Beta}(\alpha_{ij}^{(\mathcal{T})},\beta_{ij}^{(\mathcal{T})})
\right\}_{i=1,j=1}^{\mathcal{N}_s,\mathcal{N}_r}.
\end{equation}

We wish to select $t$ stimulation trials according to an objective $\mathcal{L}^{(t)}$, such that
\begin{equation}\label{eqn:kld}
\mathcal{L}^{(t)}
=
D_{\mathrm{KL}}\!\left(
\Theta^{(\mathcal{T})} \,\|\, \Theta^{(t)}
\right)
\end{equation}

is minimized, where
\begin{align}
\mathcal{L}^{(t)}
&=
\frac{1}{\mathcal{N}_s\mathcal{N}_r}
\sum_{i=1}^{\mathcal{N}_s}
\sum_{j=1}^{\mathcal{N}_r}
D_{\mathrm{KL}}
\left[
\mathrm{Beta}(\alpha_{ij}^{(\mathcal{T})},\beta_{ij}^{(\mathcal{T})})
\;\|\;
\mathrm{Beta}(\alpha_{ij}^{(t)},\beta_{ij}^{(t)})
\right]\\
{}
&=
\frac{1}{\mathcal{N}_s\mathcal{N}_r}
\sum_{i=1}^{\mathcal{N}_s}
\sum_{j=1}^{\mathcal{N}_r}
\Bigg[
\log\frac{B(\alpha_{ij}^{(t)},\beta_{ij}^{(t)})}{B(\alpha_{ij}^{(\mathcal{T})},\beta_{ij}^{(\mathcal{T})})}
+
(\alpha_{ij}^{(\mathcal{T})}-\alpha_{ij}^{(t)})\psi(\alpha_{ij}^{(\mathcal{T})})
+
(\beta_{ij}^{(\mathcal{T})}-\beta_{ij}^{(t)})\psi(\beta_{ij}^{(\mathcal{T})})
\notag \\
&\quad+
(\alpha_{ij}^{(t)}+\beta_{ij}^{(t)}-\alpha_{ij}^{(\mathcal{T})}-\beta_{ij}^{(\mathcal{T})})
\psi(\alpha_{ij}^{(\mathcal{T})}+\beta_{ij}^{(\mathcal{T})})
\Bigg]
\end{align}

and where $B(\cdot,\cdot)$ is the Beta function, $\psi(\cdot)$ is the digamma function, and $D_{\mathrm{KL}}$ is the Kullback-Leibler Divergence (KLD). This can easily be decomposed row-wise, for a stimulation target $i$: 
\begin{align}
\mathcal{L}_i^{(t)}
=
\frac{1}{\mathcal{N}_r}
\sum_{j=1}^{\mathcal{N}_r}
D_{\mathrm{KL}}
\left[
\mathrm{Beta}(\alpha_{ij}^{(\mathcal{T})},\beta_{ij}^{(\mathcal{T})})
\;\|\;
\mathrm{Beta}(\alpha_{ij}^{(t)},\beta_{ij}^{(t)})
\right]
\rightarrow
\mathcal{L}^{(t)}
=
\frac{1}{\mathcal{N}_s}
\sum_{i=1}^{\mathcal{N}_s}
\mathcal{L}_i^{(t)}.
\end{align}

The extended problem is to identify \textit{combinations} of targets to
stimulate jointly. Let \(\mathcal{V}=\{1,\dots,\mathcal{N}_s\}\) denote the
set of all stimulation targets. We define the combinatorial action space as
\begin{equation}\label{eqn:combspace}
\mathcal{C}
=
\left\{
S \subseteq \mathcal{V}
\;:\;
1 \le |S| \le k_{\max}
\right\},
\end{equation}
where \(k_{\max}\) is the maximum allowed ensemble size.

For a stimulation set \(S_t \in \mathcal{C}\) and responses
\(\{r_{S_t,j}^{(t)}\}_{j=1}^{\mathcal{N}_r}\), we wish to minimize
\(\mathcal{L}^{(t)}_{S_t}\).

\subsection{Solutions to $\mathcal{N}_s$-armed and combinatorial bandits}\label{ssec:mab}
We formulate photostimulation target selection as one of two problems: standalone selection ($\mathcal{N}_s$-armed bandits) and combinatorial optimization (combinatorial bandits). We choose to split these into separate problems due to practical experiment considerations: In the former case, groups are pre-selected, for example because of shared functional barcodes \cite{Naumann2016-mx} or functional wiring hypotheses \cite{Dal-Maschio2017-dp}. For the latter, group selection can be unbiased with respect to target ensemble size and would be useful for causal exploration of functional neural circuitry. This approach is more flexible, and can generalize beyond connectivity mapping, for example to informing brain-computer interfaces \cite{Tang2024-ku}.

For standalone selection, we represent trial outcomes as
\[
\mathcal{D}^{(t)}
=
\left\{
r_{ij}^{(\tau)}
:
i \in \mathcal{V},\;
j \in \mathcal{R},\;
\tau = 1,\dots,t
\right\},
\]
where \(\mathcal{V}=\{1,\dots,\mathcal{N}_s\}\) and \(\mathcal{R}=\{1,\dots,\mathcal{N}_r\}\)
denote stimulation targets and responders.

Since the action space only includes selection of a particular target, and since we assume no coupling between subsequent trials, we can frame action selection as $\mathcal{N}_s$ arms in a bandit problem, where each arm corresponds to target
$i \in \mathcal{V}$. Selecting arm $i$ at trial $t$ yields
\[
\mathbf{r}_i^{(t)}
=
\left(
r_{i1}^{(t)}, \dots, r_{i\mathcal{N}_r}^{(t)}
\right),
\qquad
r_{ij}^{(t)} \sim \mathrm{Bernoulli}(\theta_{ij}).
\]

We compare OPhELIA's performance on this standalone problem to other well-known multi-armed bandit solutions, including Thompson Sampling (TS) and Upper Confidence Bounds (UCB). We also consider a uniform sampling strategy, which we call Round Robin (RR) (Sec.~\ref{sssec:ts}--\ref{sssec:rr}).

For combinatorial selection, with target ensembles of maximum size
\(k_{\max}\), each action corresponds to a stimulation subset
\(S_t \in \mathcal{C}\), where $\mathcal{C}$ includes all possible target combinations. Recalling Eq. \ref{eqn:combspace}:
\[
\mathcal{C}
=
\left\{
S \subseteq \mathcal{V}
:
1 \le |S| \le k_{\max}
\right\}
\text{, where }
|\mathcal{C}|=
\sum_{m=1}^{k_{\max}}
{\mathcal{N}_s \choose m},
\]
recovering \(2^{\mathcal{N}_s}-1\) when \(k_{\max}=\mathcal{N}_s\).

Analogous to standalone selection, at trial \(t\), stimulating subset \(S_t\) yields response vector $\mathbf{r}_{S_t}$:
\[
\mathbf{r}_{S_t}^{(t)}
=
\left(
r_{S_t,1}^{(t)},
\dots,
r_{S_t,\mathcal{N}_r}^{(t)}
\right),
\qquad
r_{S_t,j}^{(t)}
\sim
\mathrm{Bernoulli}(\theta_{S_t,j}).
\]

For the combinatorial case, we compare OPhELIA (Sec. \ref{ssec:ophelia}) with representative baselines adapted for probabilistic connectivity inference, including NGT and LRM. Note that these implementations are adapted to our graded probabilistic objective and are therefore motivated by, and not reproductions of \cite{Draelos2020-sd,Draelos2020-zj, Bull2024-vh, Wagenmaker2024-active}. Details on these implementations are discussed in Sec. \ref{sssec:NGT}-\ref{sssec:LRM}.

\subsection{The OPhELIA pipeline}\label{ssec:ophelia}
OPhELIA comprises four components: (1) A Bayesian inference framework for iterative connectivity mapping (Sec. \ref{ssec:bayfconn}); (2) an action selection heuristic designed to maximize information gain with each target selection, which we call ambiguity maximization (Sec. \ref{sssec:ambigmax}); (3) the use of priors to predict which candidate target neurons and groups will be maximally informative for connectivity inference; and (4) the integration of each of (1)-(3) with key standalone and combinatorial strategies, active learning and compressed sensing. These components support sequential online updating, although the present experiments evaluate them through post-hoc replay of previously collected
trials.

\subsubsection{Ambiguity maximization as a heuristic}\label{sssec:ambigmax}
Ambiguity maximization is an adaptive heuristic for selecting informative photostimulation targets under limited trial budgets. Informative targets are those whose downstream responses remain unresolved, motivating selection based on both posterior uncertainty and proximity to maximal Bernoulli entropy. Related adaptive noisy group-testing approaches prioritize candidate neurons whose posterior binary connection probabilities are closest to $0.5$ \cite{Draelos2020-sd,Draelos2020-zj}. OPhELIA extends this principle to graded target-responder response probabilities using an acquisition score that combines Beta-posterior variance with proximity to $0.5$, aggregates across responders, and generalizes to standalone and combinatorial action spaces.

From our Beta-Bernoulli connectivity formulation (Sec. \ref{ssec:bayfconn}), we can consider the posterior variance:
\begin{equation}
\mathrm{Var}[\theta_{ij}^{(t)}] = \frac{\alpha_{ij}^{(t)} \beta_{ij}^{(t)}}{(\alpha_{ij}^{(t)} + \beta_{ij}^{(t)})^2 (\alpha_{ij}^{(t)} + \beta_{ij}^{(t)} + 1)},
\end{equation}

and construct a heuristic motivated by the entropy of each Bernoulli response. For a Bernoulli random variable with response probability $\theta$, the entropy is
\begin{equation}\label{eqn:entropy}
\mathcal{H}(\theta)
=
-\theta \log \theta - (1-\theta)\log(1-\theta).
\end{equation}

Its first and second derivatives are
\begin{equation}
\frac{d\mathcal{H}}{d\theta}
=
\log(1-\theta)-\log(\theta),
\qquad
\frac{d^2\mathcal{H}}{d\theta^2}
=
-\frac{1}{\theta} - \frac{1}{1-\theta} < 0.
\end{equation}

Setting $d\mathcal{H}/d\theta=0$ gives $\theta=0.5$, and since the second derivative is negative for $\theta\in(0,1)$, $\mathcal{H}(\theta)$ is strictly concave and maximized at $\theta=0.5$. Thus, responses with posterior means near $0.5$ are maximally uncertain and yield the highest entropy. Since we seek to maximize expected entropy reduction, we can define an ambiguity score for selecting informative targets:
\vspace{-.1cm}
\begin{equation}\label{eqn:ambig}
\mathcal{A}_{ij} = \left(1 - 2 \left| \hat{\theta}_{ij} - 0.5 \right| \right) \cdot \mathrm{Var}[\theta_{ij}], \text{ where we select } i_t = \arg\max_i \sum_{j=1}^{\mathcal N_r}\mathcal{A}_{ij}=\arg\max_i U(i).
\end{equation}

This follows equivalently for combinations $\mathcal{C}$, where we can define $\mathbf{C} \in \{0,1\}^{|\mathcal{C}| \times \mathcal{N}_s}$ as a binary design matrix, where each row $\mathbf{c}_k$ encodes a subset $C_k \in \mathcal{C}$. Then:
\begin{equation}
A_{kj}
=
\left(1 - 2\left|\hat{P}_{kj} - 0.5\right|\right)
\cdot \mathrm{Var}[\theta_{kj}] 
\rightarrow
U(C_k)
=
\sum_{j=1}^{\mathcal{N}_r} A_{kj}
\text{, where }
C_t
=
\arg\max_{C_k \in \mathcal{C}} U(C_k).
\end{equation}

\subsubsection{Prior information for action selection}\label{ssec:hybrid}
We construct prior-informed variants of Bayesian Active Learning (OPh-AL, Sec. \ref{sssec:bal}) and Compressed Sensing (OPh-CS, Sec. \ref{sssec:cs}) by learning which stimulation targets are likely to produce high downstream uncertainty before causal perturbation experiments begin.

For each stimulation target \(i\), we define its mean responder entropy as
\begin{equation}
\bar{\mathcal{H}}_i
=
\frac{1}{\mathcal{N}_r}
\sum_{j=1}^{\mathcal{N}_r}
\mathcal{H}(\theta_{ij}),
\end{equation}
and train a random forest regressor \(f(\mathbf{x}_i)\) using spontaneous and
visually evoked activity features \(\mathbf{x}_i\) to predict target-level entropy,
\(\hat{\mathcal{H}}_i=f(\mathbf{x}_i)\). We then define the prior-informed
acquisition score
\begin{equation}
U_{\mathrm{pinf}}(i)
=
(1-\lambda)U(i)
+
\lambda\hat{\mathcal{H}}_i,
\qquad
i_t^{*}
=
\arg\max_i U_{\mathrm{pinf}}(i),
\end{equation}
where we fix \(\lambda=0.35\).

Features considered were derived from data collected before photostimulation trials. See Fig. \ref{fig:experiment_results}-a and b, which shows spontaneous and evoked responses in a single target cell; and Sec. \ref{ssec:datagen} and \ref{app:rf_prior}.

\section{Results}

\subsection{Simulations}\label{ssec:sims1}
\begin{figure}
    \centering
    \includegraphics[width=1\linewidth]{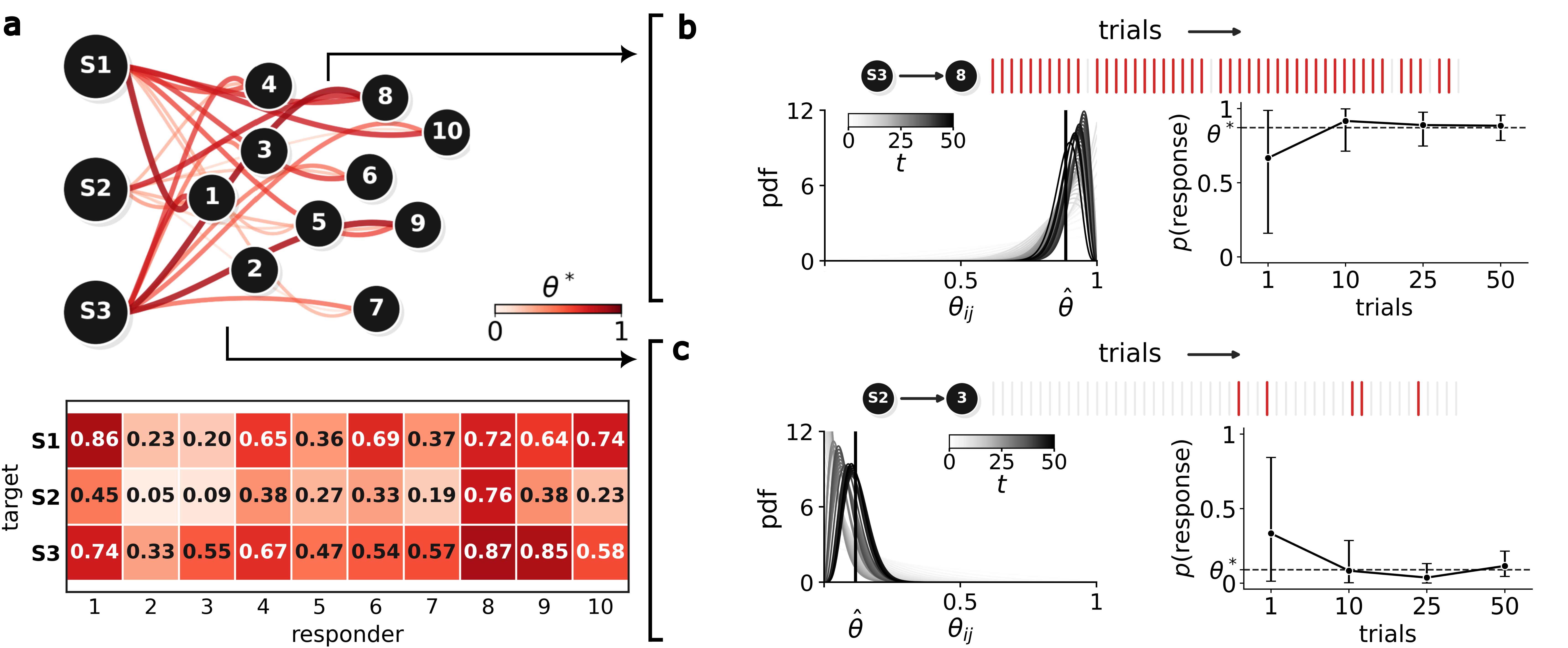}
    \caption{\textbf{Monte Carlo simulations for Bayesian connectivity inference.} \textbf{a}. Example simulated photostimulation experiment. For three candidate photostimulation targets (S1-S3), there is an underlying true probability of response ($\theta^*$) in each downstream responder. For high and low response weights (S3 to neuron 8, S2 to neuron 3), we show (counter-clockwise in \textbf{b}.,\textbf{c}.) the response outcomes, corresponding Beta distribution evolution, and closeness to $\theta^*$ over 1, 10, 25, and 50 discrete trials. Error bars represent the top and bottom 2.5\% margins of the posterior distribution.}
    \label{fig:bernoullisim}
\end{figure}

Prior to data collection, we ran Monte Carlo simulations of stimulated target cell and responder outcomes in order to contrast the efficacy of each approach on synthetic data. We created a simulation with 24 photostimulated neurons and 20 responders, activating each target for 50 independent trials. A toy illustration is shown in Fig. \ref{fig:bernoullisim} for reference, with three targets and 10 responders, and each target stimulated for 50 trials. Fig. \ref{fig:bernoullisim}-a shows known ground truth values $\theta^*$ of response probabilities for all connections. For any responder $j$ from target $i$, we infer a Beta-distributed connection probability, $\theta_{ij}$, and take the expected value as our connection estimate, $\hat{\theta}$ (Fig \ref{fig:bernoullisim}-b and c, bottom left). Fig. \ref{fig:bernoullisim}-b and c (bottom right) show that known response probabilities are recovered by our trial design; namely, that the expectation of the posterior approximately reaches the ground truth value above 10 trials.

\begin{figure}
    \centering
    \includegraphics[width=1\linewidth]{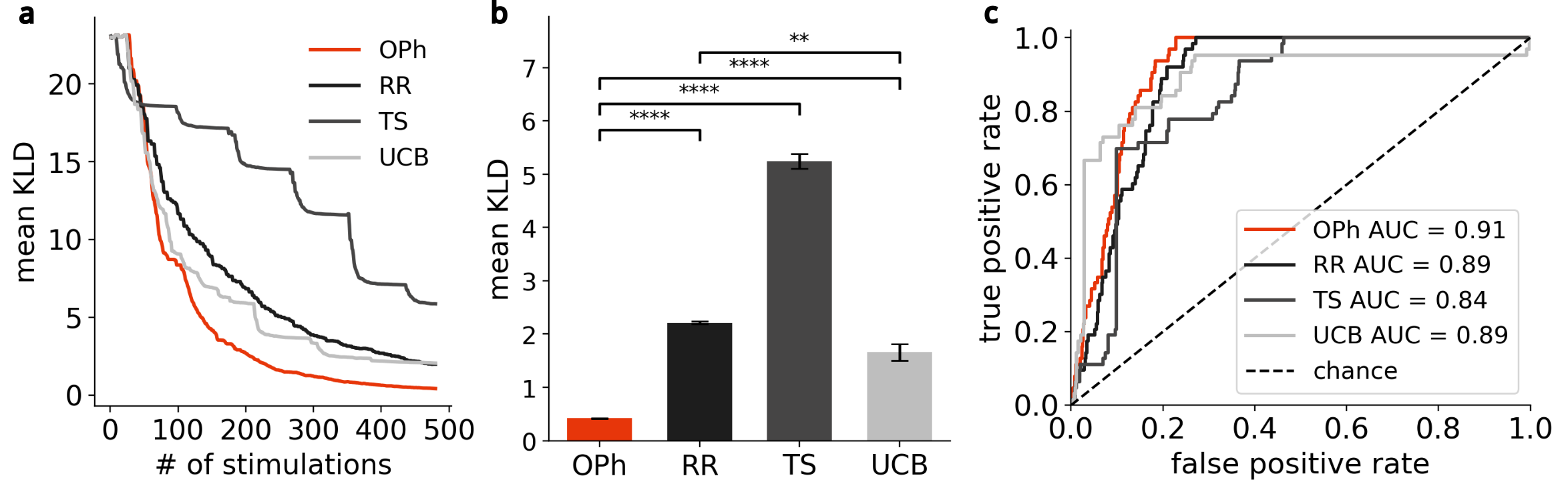}
    \caption{\textbf{An ambiguity heuristic improves connectivity estimates.} Performance of OPhELIA, with ambiguity maximization (OPh), Thompson Sampling (TS), Upper Confidence Bound (UCB), and Round Robin (RR) strategies on simulation data. \textbf{a}. For 25\% of total trials, trial-by-trial evolution of each algorithm's Kullback-Leibler divergence (KLD) to an exhaustive connectome after all trials; \textbf{b}. The same as (a), but repeated for 20 runs and using a Wilcoxon Signed-Rank Test (${}^{**}=p<0.005, {}^{****}=p<0.00005$); \textbf{c}. Receiver operating characteristic (ROC) curves for recovery of the top 5\% highest-entropy edges in $\hat{\Theta}$ as compared to chance prediction of these edges.}
    \label{fig:motion_prior}
\end{figure}

Comparing each standalone strategy (Sec. \ref{ssec:mab}) on our simulation based on the disagreement of each of their inferred functional connectomes with a trial-exhaustive connectome (Eq. \ref{eqn:kld}), we observe in Fig. \ref{fig:motion_prior}-a and b that round robin (RR) actions approach exhaustive connection estimates faster than Thompson Sampling (TS), but not Upper Confidence Bounds (UCB), which suggests that, particularly for an under-constrained bandit problem, a uniform approach can acquire informative selections more efficiently than posterior sampling. Compared to all other strategies, OPhELIA (OPh) most closely matches our trial-exhaustive network up to and at a chosen 25\% trial threshold. This threshold was empirically motivated from previous standalone experiments. In Fig. \ref{fig:motion_prior}-c, where we evaluate the ability of each strategy to predict informative (high entropy) edges, we see the previous result pattern, which in this case is consistent with the hypothesis that informative action selection is guided by maximum entropy. Comparing RR to OPh, RR by its construction under-samples high-entropy rows, spending its trial budget on rows whose connectivity distributions are already relatively well resolved.

\subsection{Experiments}

\begin{figure}[!h]
    \centering
    \includegraphics[width=1\linewidth]{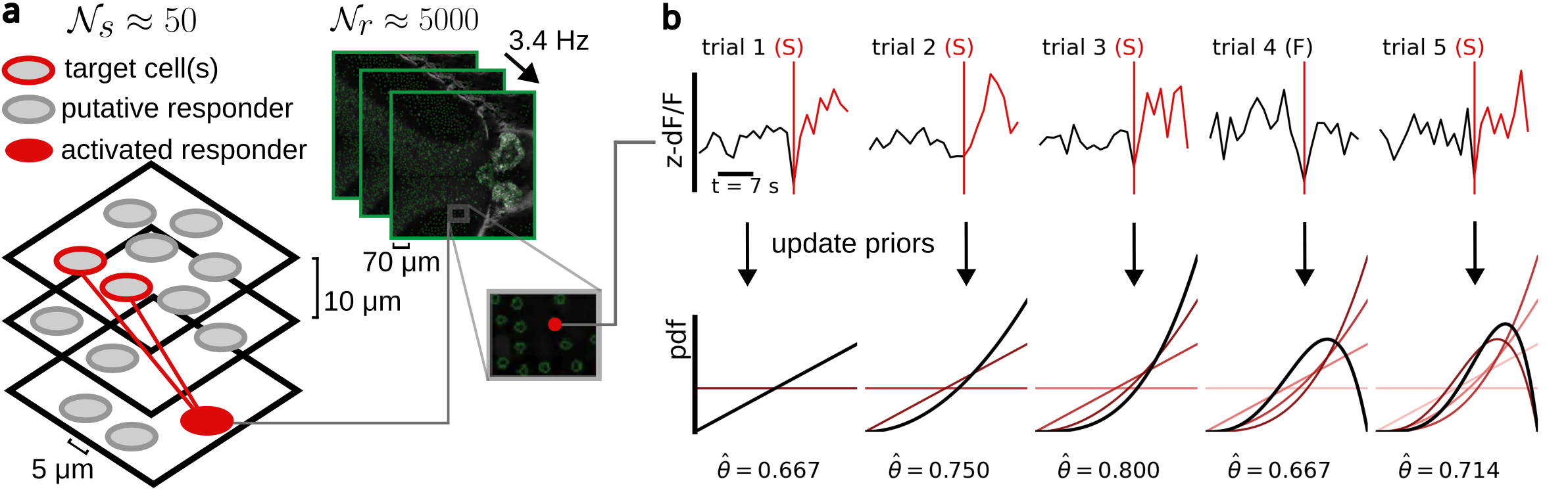}
    \caption{\textbf{Collection of experimental data for validation of simulation results.} \textbf{a.} We perform 2-photon holographic optogenetic photostimulation on a target of one or more neurons out of $\approx$ 50 candidate locations ($\mathcal{N}_s$), and collect responses across midbrain regions of interest in $\approx$ 5,000 cells ($\mathcal{N}_r$); \textbf{b.} (left to right) Causal trial traces and outcomes (S = success, F = fail) for an example stimulation target and responder. Corresponding posteriors are shown below trial outcomes (darker shades indicate more recent trials). See Sec. \ref{sssec:exp_criteria} for more on our S and F decision criteria.}
    \label{fig:experiment_data}
\end{figure}

We next applied standalone strategies to real experimental photostimulation and calcium imaging data. Fig. \ref{fig:experiment_data} illustrates the broad probabilistic connectivity estimation pipeline on this data. Fig. \ref{fig:experiment_data}-a illustrates multiple target cells being stimulated simultaneously, with a responder's activity measured in a separate imaging plane. Putative target distances are at least 5 $\mu m$ apart. Fig. \ref{fig:experiment_data}-b (top) shows, for the same target and responder, trial response outcomes as $\{S,F\}$. Trace outcomes update an underlying Beta distribution, from $\mathrm{Beta}(1,1)$, as shown in Fig. \ref{fig:experiment_data}-b (bottom). 

\begin{figure}[!h]
    \centering
    \includegraphics[width=1\linewidth]{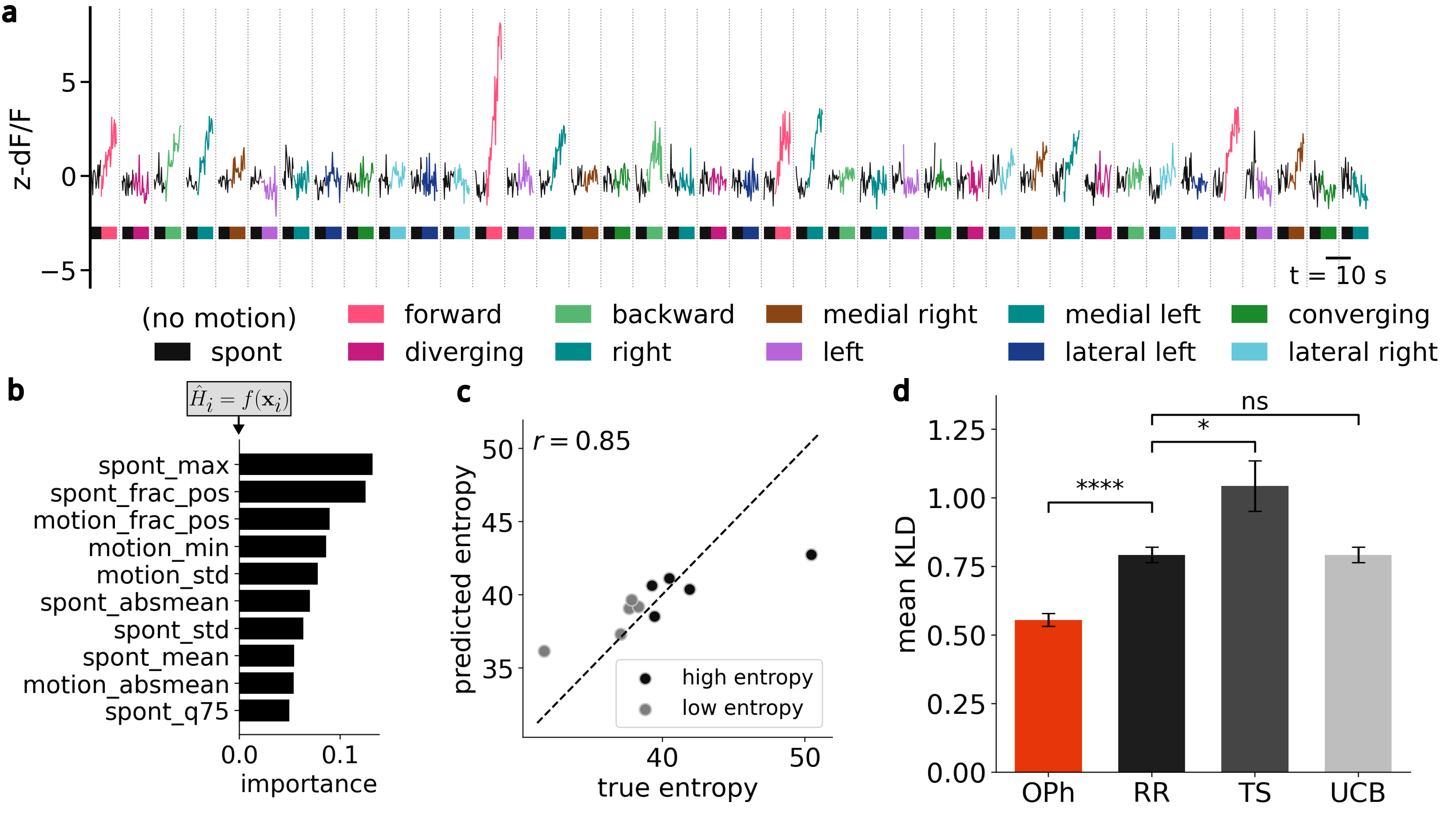}
    \caption{\textbf{Experimental priors improve inferred connectivity estimates}. \textbf{a}. Concatenated spontaneous and motion-responsive z-scored normalized fluorescence for a sample neuron, with each stimulus sorted by type; \textbf{b}. We employ regressor $f$ on feature vector $\mathbf{x}_i$ to predict mean entropy of causal connectivity mappings for stimulated targets $i$. Features are drawn from the spontaneous and motion-driven priors. We observe that the maximum value of the spontaneous curve, and the proportion of non-zero z-scored values, helps to determine the predicted entropy. \textbf{c}. Predicted versus true entropy scores for the RF regressor, showing that prior features inform separable output classes (high and low entropy targets); \textbf{d}. Using motion features to predict entropy from stimulation decisions (OPh), we observe a significant improvement in performance using 25\% of total trials. Significance was measured by a Wilcoxon Signed-Rank Test (${}^{*}=p<0.05, {}^{****}=p<0.00005$).}
    \label{fig:experiment_results}
\end{figure}

As mentioned in Sec. \ref{ssec:hybrid}, we used an RF regressor (Fig. \ref{fig:experiment_results}-b), and identified features from $\mathbf{x}_i$ for stimulation target $i$ which predict mean entropy in downstream responders. Z-scored traces (spontaneous and motion-evoked) are shown in Fig. \ref{fig:experiment_results}-a. Building on our ambiguity heuristic, we incorporated these features into OPhELIA as priors. In Fig. \ref{fig:experiment_results}-c, it is suggested by the strong predicted and true correlations, that prior physiological neural activity can predict which stimulation targets will result in high-entropy or low-entropy downstream connectivity probabilities \cite{Bounds2025-ba}. This is validated in Fig. \ref{fig:experiment_results}-d, where we find that compared to strategies which do not use prior information, OPhELIA (OPh) achieves a significantly lower mean KL divergence (KLD) to an exhaustive trial connectome. 

\begin{figure}[!h]
    \centering
    \includegraphics[width=1.03\linewidth]{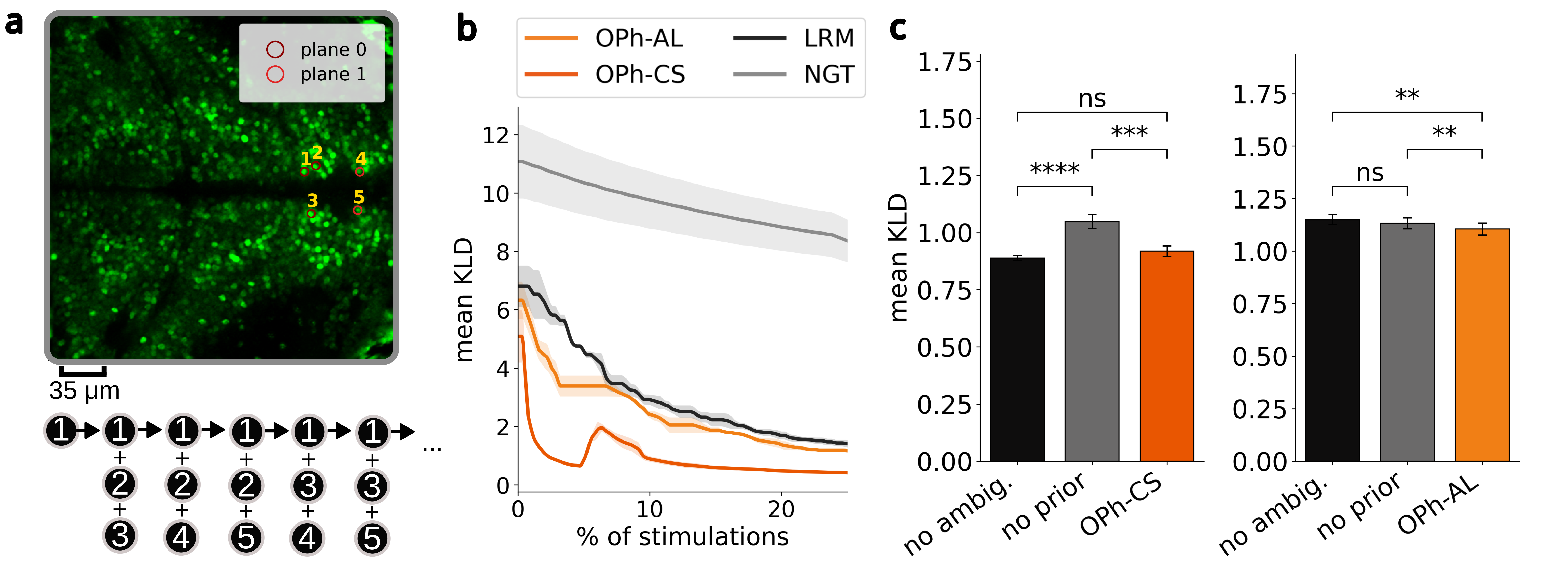}
    \caption{\textbf{OPhELIA infers trial-exhaustive connectivity using 5\% of trials.} \textbf{a.} Plane-collapsed image showing target selection (above) and a subset of selected cells (below); \textbf{b}. Mean KL divergence (KLD) between exhaustive photostimulation of targets using NGT, LRM, and OPhELIA with Active Learning (OPh-AL) and Compressed Sensing (OPh-CS) across three subjects ($\mathcal{N}_r$=500). Shaded regions represent the standard error of the mean (SEM); \textbf{c.} Ablations of the ambiguity maximization heuristic (no ambig.) and prior component (no prior) after 25\% of trials and over 30 algorithm runs, 10 per fish (Wilcoxon Signed-Rank Test, ${}^{**}=p<0.005$, ${}^{***}=p<0.0005$,  ${}^{****}=p<0.00005$).}
    \label{fig:combo}
\end{figure}

Finally, we considered combinatorial optimization across 21 experimentally measured target ensembles formed from $\mathcal{N}_s=5$ candidate targets in each of three fish, excluding all two-target ensembles (Sec. \ref{sssec:exp_comb_stim}). For each fish, we analyzed a fixed subset of $\mathcal{N}_r=500$ responders, with 12 trials per ensemble retained for analysis. We compared the trial efficiency and connectome-inference performance (KLD) of OPh-AL and OPh-CS with our noisy group testing (NGT) and low-rank matrix recovery (LRM) implementations. All methods used an ambiguity-based acquisition policy (Sec. \ref{sssec:NGT}, \ref{sssec:LRM}). \textit{In vivo} data was collected before algorithm comparison in separate spontaneous and photostimulation windows. Figure \ref{fig:combo}-a shows the selected targets and a subset of tested combinations for a representative fish. Across the evaluated trial budgets, OPh-AL achieved lower mean KLD than the adapted LRM and NGT implementations, while OPh-CS most closely approximated the trial-exhaustive reference, achieving the lowest mean KLD at budgets as low as 5\% of the available trials (Fig. \ref{fig:combo}-b). Ablation analyses at 25\% of the trial budget showed that removing prior information degraded performance for OPh-CS, whereas removing either the ambiguity heuristic or prior increased KLD for OPh-AL (Fig. \ref{fig:combo}-c). Together, these suggest that prior information is a key aspect of OPh-CS, and that both the ambiguity heuristic and priors inform the performance of OPh-AL.

\section{Conclusion}\label{sec:conclusion}
We presented OPhELIA, an adaptive Bayesian framework for causal connectome inference under limited trial budgets. In simulations, ambiguity-guided selection improved standalone mapping relative to Round Robin, Thompson Sampling, and Upper Confidence Bounds. In \textit{in vivo} experiments, spontaneous and visually evoked activity predicted target informativeness, supporting the use of learned physiological priors for action selection. In the combinatorial setting, OPh-CS most closely approximated the trial-exhaustive reference using 5\% of the available trials, while ablations revealed distinct contributions of prior information and ambiguity maximization across active learning and compressed sensing. These results demonstrate that OPhELIA can improve the trial efficiency of causal circuit mapping whilst achieving strong alignment to the underlying exhaustive connectome.

\paragraph{Limitations}
Our approach assumes that a photostimulated neuron is successfully activated. We therefore do not optimize for activation of the stimulation site using experimental covariates such as laser power \cite{Triplett2023-pr}. Incorporating online calibration of stimulation quality would be a valuable extension. Furthermore, our Beta-Bernoulli model, separate from regression strategies \cite{Bull2024-vh}, reduces responses to binary evoked or non-evoked outcomes, which provide robustness, interpretability, and efficient uncertainty updates, but discard information about graded response amplitudes and temporal structure, and thus would not be appropriate to investigate, for example, responder dynamics. 

Looking forward, OPhELIA may improve other causal perturbation strategies where interventions are time-constrained, for example in focused ultrasound neuromodulation \cite{Murphy2024-ra}. We furthermore anticipate that future iterations will incorporate online deployment during live experiments, for example using a platform such as improv \cite{Draelos2021-sb}; and hope that this and following projects will help to establish adaptive intervention learning as a practical paradigm for next-generation circuit neuroscience.

\newpage
\bibliographystyle{ieeetr}
\bibliography{neurips_2026}

@ARTICLE{Triplett2022-ba,
  title    = "Rapid learning of neural circuitry from holographic ensemble
              stimulation enabled by model-based compressed sensing",
  author   = "Triplett, Marcus A and Gajowa, Marta and Antin, Benjamin and
              Sadahiro, Masato and Adesnik, Hillel and Paninski, Liam",
  journal  = "bioRxiv",
  pages    = "2022.09.14.507926",
  month    =  sep,
  year     =  2022,
  language = "en"
}

@ARTICLE{Pegard2017-tb,
  title     = "Three-dimensional scanless holographic optogenetics with temporal
               focusing ({3D}-{SHOT})",
  author    = "Pégard, Nicolas C and Mardinly, Alan R and Oldenburg, Ian Antón
               and Sridharan, Savitha and Waller, Laura and Adesnik, Hillel",
  journal   = "Nat. Commun.",
  publisher = "Nature Publishing Group",
  volume    =  8,
  number    =  1,
  pages     =  1228,
  month     =  oct,
  year      =  2017,
  language  = "en"
}

@ARTICLE{Draelos2020-sd,
  title   = "Online Neural Connectivity Estimation with Noisy Group Testing",
  author  = "Draelos, Anne and Pearson, John",
  journal = "Advances in Neural Information Processing Systems",
  volume  =  33,
  pages   = "7437--7448",
  year    =  2020
}

@ARTICLE{Weiler2024-pk,
  title     = "Overcoming off-target optical stimulation-evoked cortical
               activity in the mouse brain in vivo",
  author    = "Weiler, Simon and Velez-Fort, Mateo and Margrie, Troy W",
  journal   = "iScience",
  publisher = "Elsevier BV",
  volume    =  27,
  number    =  11,
  pages     =  111152,
  month     =  nov,
  year      =  2024,
  language  = "en"
}

@ARTICLE{Triplett2023-pr,
  title     = "Bayesian target optimisation for high-precision holographic
               optogenetics",
  author    = "Triplett, M and Gajowa, Marta A and Adesnik, H and Paninski, L",
  editor    = "Oh, A and Naumann, T and Globerson, A and Saenko, K and Hardt, M
               and Levine, S",
  journal   = "bioRxiv",
  publisher = "Curran Associates, Inc.",
  volume    =  36,
  pages     = "10972--10994",
  month     =  may,
  year      =  2023
}

@ARTICLE{Draelos2021-sb,
  title       = "\textit{improv}: A software platform for real-time and adaptive
                 neuroscience experiments",
  author      = "Draelos, Anne and Loring, Matthew D and Nikitchenko, Maxim and
                 Sriworarat, Chaichontat and Gupta, Pranjal and Sprague, Daniel
                 Y and Pnevmatikakis, Eftychios and Giovannucci, Andrea and
                 Benster, Tyler and Deisseroth, Karl and Pearson, John M and
                 Naumann, Eva A",
  journal     = "bioRxiv",
  institution = "bioRxiv",
  pages       = "2021.02.22.432006",
  month       =  feb,
  year        =  2021,
  language    = "en"
}

@ARTICLE{Giovannucci2019-ix,
  title     = "{CaImAn} an open source tool for scalable calcium imaging data
               analysis",
  author    = "Giovannucci, Andrea and Friedrich, Johannes and Gunn, Pat and
               Kalfon, Jérémie and Brown, Brandon L and Koay, Sue Ann and
               Taxidis, Jiannis and Najafi, Farzaneh and Gauthier, Jeffrey L and
               Zhou, Pengcheng and Khakh, Baljit S and Tank, David W and
               Chklovskii, Dmitri B and Pnevmatikakis, Eftychios A",
  journal   = "Elife",
  publisher = "eLife Sciences Publications, Ltd",
  volume    =  8,
  month     =  jan,
  year      =  2019,
  language  = "en"
}

@ARTICLE{Pachitariu2016-vp,
  title       = "{Suite2p}: beyond 10,000 neurons with standard two-photon
                 microscopy",
  author      = "Pachitariu, Marius and Stringer, Carsen and Dipoppa, Mario and
                 Schröder, Sylvia and Rossi, L Federico and Dalgleish, Henry and
                 Carandini, Matteo and Harris, Kenneth D",
  journal     = "bioRxiv",
  institution = "bioRxiv",
  pages       =  061507,
  month       =  jun,
  year        =  2016,
  language    = "en"
}

@ARTICLE{Naumann2016-mx,
  title     = "From whole-brain data to functional circuit models: The zebrafish
               optomotor response",
  author    = "Naumann, Eva A and Fitzgerald, James E and Dunn, Timothy W and
               Rihel, Jason and Sompolinsky, Haim and Engert, Florian",
  journal   = "Cell",
  publisher = "Elsevier",
  volume    =  167,
  number    =  4,
  pages     = "947--960.e20",
  month     =  nov,
  year      =  2016,
  language  = "en"
}

@ARTICLE{Bounds2025-ba,
  title     = "Network influence determines the impact of cortical ensembles on
               stimulus detection",
  author    = "Bounds, Hayley A and Adesnik, Hillel",
  journal   = "Neuron",
  publisher = "Elsevier BV",
  volume    =  113,
  number    =  14,
  pages     = "2358--2369.e5",
  month     =  jul,
  year      =  2025,
  language  = "en"
}

@ARTICLE{Chen2025-pu,
  title     = "High-throughput synaptic connectivity mapping using in vivo
               two-photon holographic optogenetics and compressive sensing",
  author    = "Chen, I-Wen and Chan, Chung Yuen and Navarro, Phillip and de
               Sars, Vincent and Ronzitti, Emiliano and Oweiss, Karim and
               Tanese, Dimitrii and Emiliani, Valentina",
  journal   = "Nat. Neurosci.",
  publisher = "Springer Science and Business Media LLC",
  volume    =  28,
  number    =  10,
  pages     = "2141--2153",
  month     =  oct,
  year      =  2025,
  language  = "en"
}

@ARTICLE{Triplett2026-ut,
  title     = "Computational optimization of two-photon holographic stimulation
               sitesin vivo",
  author    = "Triplett, Marcus A and Bäumler, Edgar and Prodan, Alex and
               Stonis, Rokas and Peterka, Darcy S and Häusser, Michael and
               Paninski, Liam",
  journal   = "J. Neural Eng.",
  publisher = "IOP Publishing",
  volume    =  23,
  number    =  2,
  pages     =  026015,
  month     =  mar,
  year      =  2026,
  language  = "en"
}

@ARTICLE{Svara2022-ik,
  title     = "Automated synapse-level reconstruction of neural circuits in the
               larval zebrafish brain",
  author    = "Svara, Fabian and Förster, Dominique and Kubo, Fumi and
               Januszewski, Michał and Dal Maschio, Marco and Schubert, Philipp
               J and Kornfeld, Jörgen and Wanner, Adrian A and Laurell, Eva and
               Denk, Winfried and Baier, Herwig",
  journal   = "Nat. Methods",
  publisher = "Springer Science and Business Media LLC",
  volume    =  19,
  number    =  11,
  pages     = "1357--1366",
  month     =  nov,
  year      =  2022,
  language  = "en"
}

@ARTICLE{Wolf2017-nh,
  title     = "Sensorimotor computation underlying phototaxis in zebrafish",
  author    = "Wolf, Sébastien and Dubreuil, Alexis M and Bertoni, Tommaso and
               Böhm, Urs Lucas and Bormuth, Volker and Candelier, Raphaël and
               Karpenko, Sophia and Hildebrand, David G C and Bianco, Isaac H
               and Monasson, Rémi and Debrégeas, Georges",
  journal   = "Nat. Commun.",
  publisher = "Springer Science and Business Media LLC",
  volume    =  8,
  number    =  1,
  pages     =  651,
  month     =  sep,
  year      =  2017,
  language  = "en"
}

@ARTICLE{Fouke2025-xy,
  title     = "Neural circuits underlying divergent visuomotor strategies of
               zebrafish and Danionella cerebrum",
  author    = "Fouke, Kaitlyn E and He, Zichen and Loring, Matthew D and
               Naumann, Eva A",
  journal   = "Curr. Biol.",
  publisher = "Elsevier BV",
  month     =  may,
  year      =  2025,
  language  = "en"
}

@ARTICLE{Emiliani2015-uk,
  title     = "All-optical interrogation of neural circuits",
  author    = "Emiliani, Valentina and Cohen, Adam E and Deisseroth, Karl and
               Häusser, Michael",
  journal   = "J. Neurosci.",
  publisher = "Society for Neuroscience",
  volume    =  35,
  number    =  41,
  pages     = "13917--13926",
  month     =  oct,
  year      =  2015,
  language  = "en"
}

@ARTICLE{Yang2018-ey,
  title     = "Simultaneous two-photon imaging and two-photon optogenetics of
               cortical circuits in three dimensions",
  author    = "Yang, Weijian and Carrillo-Reid, Luis and Bando, Yuki and
               Peterka, Darcy S and Yuste, Rafael",
  journal   = "Elife",
  publisher = "eLife Sciences Publications, Ltd",
  volume    =  7,
  number    = "e32671",
  month     =  feb,
  year      =  2018,
  language  = "en"
}

@ARTICLE{Marshel2019-yp,
  title     = "Cortical layer-specific critical dynamics triggering perception",
  author    = "Marshel, James H and Kim, Yoon Seok and Machado, Timothy A and
               Quirin, Sean and Benson, Brandon and Kadmon, Jonathan and Raja,
               Cephra and Chibukhchyan, Adelaida and Ramakrishnan, Charu and
               Inoue, Masatoshi and Shane, Janelle C and McKnight, Douglas J and
               Yoshizawa, Susumu and Kato, Hideaki E and Ganguli, Surya and
               Deisseroth, Karl",
  journal   = "Science",
  publisher = "American Association for the Advancement of Science (AAAS)",
  volume    =  365,
  number    =  6453,
  pages     = "eaaw5202",
  month     =  aug,
  year      =  2019,
  language  = "en"
}

@ARTICLE{Russell2024-ad,
  title     = "The influence of cortical activity on perception depends on
               behavioral state and sensory context",
  author    = "Russell, Lloyd E and Fişek, Mehmet and Yang, Zidan and Tan, Lynn
               Pei and Packer, Adam M and Dalgleish, Henry W P and Chettih,
               Selmaan N and Harvey, Christopher D and Häusser, Michael",
  journal   = "Nat. Commun.",
  publisher = "Nature Publishing Group",
  volume    =  15,
  number    =  1,
  pages     =  2456,
  month     =  mar,
  year      =  2024,
  language  = "en"
}

@ARTICLE{Rowland2023-zx,
  title     = "Propagation of activity through the cortical hierarchy and
               perception are determined by neural variability",
  author    = "Rowland, James M and van der Plas, Thijs L and Loidolt, Matthias
               and Lees, Robert M and Keeling, Joshua and Dehning, Jonas and
               Akam, Thomas and Priesemann, Viola and Packer, Adam M",
  journal   = "Nat. Neurosci.",
  publisher = "Nature Publishing Group",
  volume    =  26,
  number    =  9,
  pages     = "1584--1594",
  month     =  sep,
  year      =  2023,
  language  = "en"
}

@ARTICLE{Vishwanathan2024-uv,
  title     = "Predicting modular functions and neural coding of behavior from a
               synaptic wiring diagram",
  author    = "Vishwanathan, Ashwin and Sood, Alex and Wu, Jingpeng and Ramirez,
               Alexandro D and Yang, Runzhe and Kemnitz, Nico and Ih, Dodam and
               Turner, Nicholas and Lee, Kisuk and Tartavull, Ignacio and
               Silversmith, William M and Jordan, Chris S and David, Celia and
               Bland, Doug and Sterling, Amy and Seung, H Sebastian and Goldman,
               Mark S and Aksay, Emre R F and {Eyewirers}",
  journal   = "Nat. Neurosci.",
  publisher = "Springer Science and Business Media LLC",
  volume    =  27,
  number    =  12,
  pages     = "2443--2454",
  month     =  dec,
  year      =  2024,
  language  = "en"
}

@ARTICLE{Houlsby2011-my,
  title         = "Bayesian active learning for classification and preference
                   learning",
  author        = "Houlsby, Neil and Huszár, Ferenc and Ghahramani, Zoubin and
                   Lengyel, Máté",
  journal       = "arXiv [stat.ML]",
  month         =  dec,
  year          =  2011,
  archivePrefix = "arXiv",
  primaryClass  = "stat.ML"
}

@ARTICLE{Werner2025-kd,
  title         = "Bayesian active Learning by distribution disagreement",
  author        = "Werner, Thorben and Schmidt-Thieme, Lars",
  journal       = "arXiv [cs.LG]",
  month         =  jan,
  year          =  2025,
  archivePrefix = "arXiv",
  primaryClass  = "cs.LG"
}

@ARTICLE{Picot2018-cj,
  title     = "Temperature rise under two-photon optogenetic brain stimulation",
  author    = "Picot, Alexis and Dominguez, Soledad and Liu, Chang and Chen,
               I-Wen and Tanese, Dimitrii and Ronzitti, Emiliano and Berto,
               Pascal and Papagiakoumou, Eirini and Oron, Dan and Tessier,
               Gilles and Forget, Benoît C and Emiliani, Valentina",
  journal   = "Cell Rep.",
  publisher = "Elsevier BV",
  volume    =  24,
  number    =  5,
  pages     = "1243--1253.e5",
  month     =  jul,
  year      =  2018,
  language  = "en"
}

@ARTICLE{Papaioannou2022-rp,
  title     = "Advantages, pitfalls, and developments of all optical
               interrogation strategies of microcircuits in vivo",
  author    = "Papaioannou, Stylianos and Medini, Paolo",
  journal   = "Front. Neurosci.",
  publisher = "Frontiers Media SA",
  volume    =  16,
  pages     =  859803,
  month     =  jun,
  year      =  2022,
  language  = "en"
}

@ARTICLE{Forli2021-hb,
  title     = "Optogenetic strategies for high-efficiency all-optical
               interrogation using blue-light-sensitive opsins",
  author    = "Forli, Angelo and Pisoni, Matteo and Printz, Yoav and Yizhar,
               Ofer and Fellin, Tommaso",
  journal   = "Elife",
  publisher = "eLife Sciences Publications, Ltd",
  volume    =  10,
  month     =  may,
  year      =  2021,
  language  = "en"
}

@ARTICLE{Navarro2023-wa,
  title     = "Compressive sensing of functional connectivity maps from
               patterned optogenetic stimulation of neuronal ensembles",
  author    = "Navarro, Phillip and Oweiss, Karim",
  journal   = "Patterns (N. Y.)",
  publisher = "Elsevier BV",
  volume    =  4,
  number    =  10,
  pages     =  100845,
  month     =  oct,
  year      =  2023,
  language  = "en"
}

@ARTICLE{Mishchenko2012-uv,
  title     = "A Bayesian compressed-sensing approach for reconstructing neural
               connectivity from subsampled anatomical data",
  author    = "Mishchenko, Yuriy and Paninski, Liam",
  journal   = "J. Comput. Neurosci.",
  publisher = "Springer Science and Business Media LLC",
  volume    =  33,
  number    =  2,
  pages     = "371--388",
  month     =  oct,
  year      =  2012,
  language  = "en"
}

@article{Stringer2019-xy,
  title={High-dimensional geometry of population responses in visual cortex},
  author={Stringer, Carsen et al.},
  journal={Nature},
  year={2019}
}

@INPROCEEDINGS{Bull2024-vh,
  title     = "Active learning of neural population dynamics using two-photon
               holographic optogenetics",
  author    = "Bull, Matthew and Daie, Kayvon and Golub, Matthew and Jamieson,
               Kevin and Mi, Lu and Rozsa, Marton and Svoboda, Karel and
               Wagenmaker, Andrew",
  editor    = "Globerson, A and Mackey, L and Belgrave, D and Fan, A and Paquet,
               U and Tomczak, J and Zhang, C",
  booktitle = "Advances in Neural Information Processing Systems 37",
  publisher = "Neural Information Processing Systems Foundation, Inc. (NeurIPS)",
  address   = "San Diego, California, USA",
  volume    =  37,
  pages     = "31659--31687",
  year      =  2024
}

@ARTICLE{Candes2006-jn,
  title     = "Robust uncertainty principles: exact signal reconstruction from
               highly incomplete frequency information",
  author    = "Candes, E J and Romberg, J and Tao, T",
  journal   = "IEEE Trans. Inf. Theory",
  publisher = "Institute of Electrical and Electronics Engineers (IEEE)",
  volume    =  52,
  number    =  2,
  pages     = "489--509",
  month     =  feb,
  year      =  2006,
  language  = "en"
}

@ARTICLE{Park2000-bm,
  title     = "Analysis of upstream elements in the {HuC} promoter leads to the
               establishment of transgenic zebrafish with fluorescent neurons",
  author    = "Park, H C and Kim, C H and Bae, Y K and Yeo, S Y and Kim, S H and
               Hong, S K and Shin, J and Yoo, K W and Hibi, M and Hirano, T and
               Miki, N and Chitnis, A B and Huh, T L",
  journal   = "Dev. Biol.",
  publisher = "Elsevier BV",
  volume    =  227,
  number    =  2,
  pages     = "279--293",
  month     =  nov,
  year      =  2000,
  language  = "en"
}

@ARTICLE{Khuansuwan2019-qb,
  title     = "A novel transgenic zebrafish line allows for in vivo
               quantification of autophagic activity in neurons",
  author    = "Khuansuwan, Sataree and Barnhill, Lisa M and Cheng, Sizhu and
               Bronstein, Jeff M",
  journal   = "Autophagy",
  publisher = "Informa UK Limited",
  volume    =  15,
  number    =  8,
  pages     = "1322--1332",
  month     =  aug,
  year      =  2019,
  language  = "en"
}

@ARTICLE{Murphy2024-ra,
  title     = "Optimized ultrasound neuromodulation for non-invasive control of
               behavior and physiology",
  author    = "Murphy, Keith R and Farrell, Jordan S and Bendig, Jonas and
               Mitra, Anish and Luff, Charlotte and Stelzer, Ina A and
               Yamaguchi, Hiroshi and Angelakos, Christopher C and Choi, Mihyun
               and Bian, Wenjie and DiIanni, Tommaso and Pujol, Esther Martinez
               and Matosevich, Noa and Airan, Raag and Gaudillière, Brice and
               Konofagou, Elisa E and Butts-Pauly, Kim and Soltesz, Ivan and de
               Lecea, Luis",
  journal   = "Neuron",
  publisher = "Elsevier BV",
  volume    =  112,
  number    =  19,
  pages     = "3252--3266.e5",
  month     =  oct,
  year      =  2024,
  language  = "en"
}

@ARTICLE{Hu2012-ct,
  title         = "Reconstruction of sparse circuits using multi-neuronal
                   excitation ({RESCUME})",
  author        = "Hu, Tao and Chklovskii, Dmitri B",
  journal       = "arXiv [q-bio.NC]",
  month         =  oct,
  year          =  2012,
  archivePrefix = "arXiv",
  primaryClass  = "q-bio.NC"
}

@ARTICLE{Lee2019-nj,
  title     = "A compressed sensing framework for efficient dissection of neural
               circuits",
  author    = "Lee, Jeffrey B and Yonar, Abdullah and Hallacy, Timothy and Shen,
               Ching-Han and Milloz, Josselin and Srinivasan, Jagan and Kocabas,
               Askin and Ramanathan, Sharad",
  journal   = "Nat. Methods",
  publisher = "Springer Science and Business Media LLC",
  volume    =  16,
  number    =  1,
  pages     = "126--133",
  month     =  jan,
  year      =  2019,
  language  = "en"
}

@ARTICLE{Draelos2025-we,
  title     = "A software platform for real-time and adaptive neuroscience
               experiments",
  author    = "Draelos, Anne and Loring, Matthew D and Nikitchenko, Maxim and
               Sriworarat, Chaichontat and Gupta, Pranjal and Sprague, Daniel Y
               and Pnevmatikakis, Eftychios and Giovannucci, Andrea and Benster,
               Tyler and Deisseroth, Karl and Pearson, John M and Naumann, Eva A",
  journal   = "Nat. Commun.",
  publisher = "Springer Science and Business Media LLC",
  volume    =  16,
  number    =  1,
  pages     =  9909,
  month     =  nov,
  year      =  2025,
  language  = "en"
}

@ARTICLE{Kishi2022-mi,
  title     = "Structural basis for channel conduction in the pump-like
               channelrhodopsin {ChRmine}",
  author    = "Kishi, Koichiro E and Kim, Yoon Seok and Fukuda, Masahiro and
               Inoue, Masatoshi and Kusakizako, Tsukasa and Wang, Peter Y and
               Ramakrishnan, Charu and Byrne, Eamon F X and Thadhani, Elina and
               Paggi, Joseph M and Matsui, Toshiki E and Yamashita, Keitaro and
               Nagata, Takashi and Konno, Masae and Quirin, Sean and Lo, Maisie
               and Benster, Tyler and Uemura, Tomoko and Liu, Kehong and
               Shibata, Mikihiro and Nomura, Norimichi and Iwata, So and Nureki,
               Osamu and Dror, Ron O and Inoue, Keiichi and Deisseroth, Karl and
               Kato, Hideaki E",
  journal   = "Cell",
  publisher = "Elsevier BV",
  volume    =  185,
  number    =  4,
  pages     = "672--689.e23",
  month     =  feb,
  year      =  2022,
  language  = "en"
}

@ARTICLE{Chen2013-nb,
  title     = "Ultrasensitive fluorescent proteins for imaging neuronal activity",
  author    = "Chen, Tsai-Wen and Wardill, Trevor J and Sun, Yi and Pulver,
               Stefan R and Renninger, Sabine L and Baohan, Amy and Schreiter,
               Eric R and Kerr, Rex A and Orger, Michael B and Jayaraman, Vivek
               and Looger, Loren L and Svoboda, Karel and Kim, Douglas S",
  journal   = "Nature",
  publisher = "Springer Science and Business Media LLC",
  volume    =  499,
  number    =  7458,
  pages     = "295--300",
  month     =  jul,
  year      =  2013,
  language  = "en"
}

@ARTICLE{Pnevmatikakis2017-qu,
  title     = "{NoRMCorre}: An online algorithm for piecewise rigid motion
               correction of calcium imaging data",
  author    = "Pnevmatikakis, Eftychios A and Giovannucci, Andrea",
  journal   = "J. Neurosci. Methods",
  publisher = "Elsevier BV",
  volume    =  291,
  pages     = "83--94",
  month     =  nov,
  year      =  2017,
  language  = "en"
}

@ARTICLE{Dal-Maschio2017-dp,
  title     = "Linking neurons to network function and behavior by two-photon
               holographic optogenetics and volumetric imaging",
  author    = "Dal Maschio, Marco and Donovan, Joseph C and Helmbrecht, Thomas O
               and Baier, Herwig",
  journal   = "Neuron",
  publisher = "Elsevier BV",
  volume    =  94,
  number    =  4,
  pages     = "774--789.e5",
  month     =  may,
  year      =  2017,
  language  = "en"
}

@ARTICLE{Chettih2019-vb,
  title     = "Single-neuron perturbations reveal feature-specific competition
               in {V1}",
  author    = "Chettih, Selmaan N and Harvey, Christopher D",
  journal   = "Nature",
  publisher = "Springer Science and Business Media LLC",
  volume    =  567,
  number    =  7748,
  pages     = "334--340",
  month     =  mar,
  year      =  2019,
  language  = "en"
}

@INCOLLECTION{Oldenburg2023-ki,
  title     = "High-speed all-optical neural interfaces with {3D} temporally
               focused holography",
  author    = "Oldenburg, Ian Antón and Bounds, Hayley Anne and Pégard, Nicolas
               C",
  booktitle = "Neuromethods",
  publisher = "Springer US",
  address   = "New York, NY",
  pages     = "101--135",
  series    = "Neuromethods",
  year      =  2023,
  language  = "en"
}

@ARTICLE{Draelos2020-zj,
  title         = "Online neural connectivity estimation with ensemble
                   stimulation",
  author        = "Draelos, Anne and Naumann, Eva A and Pearson, John M",
  journal       = "arXiv [cs.LG]",
  month         =  jul,
  year          =  2020,
  archivePrefix = "arXiv",
  primaryClass  = "cs.LG"
}

@ARTICLE{Petkova2025-ob,
  title       = "A connectomic resource for neural cataloguing and circuit
                 dissection of the larval zebrafish brain",
  author      = "Petkova, Mariela D and Januszewski, Michał and Blakely, Tim and
                 Herrera, Kristian J and Schuhknecht, Gregor F P and Tiller,
                 Robert and Choi, Jinhan and Schalek, Richard L and
                 Boulanger-Weill, Jonathan and Peleg, Adi and Wu, Yuelong and
                 Wang, Shuohong and Troidl, Jakob and Vohra, Sumit Kumar and
                 Wei, Donglai and Lin, Zudi and Bahl, Armin and Tapia, Juan
                 Carlos and Iyer, Nirmala and Miller, Zachary T and Hebert,
                 Kathryn B and Pavarino, Elisa C and Taylor, Milo and Deng,
                 Zixuan and Stingl, Moritz and Hockling, Dana and Hebling, Alina
                 and Wang, Ruohong C and Zhang, Lauren L and Dvorak, Sam and
                 Faik, Zainab and King, Jr, Kareem I and Goel, Pallavi and
                 Wagner-Carena, Julian and Aley, David and Chalyshkan, Selimzhan
                 and Contreas, Dominick and Li, Xiong and Muthukumar, Akila V
                 and Vernaglia, Marina S and Carrasco, Teodoro Tapia and
                 Melnychuck, Sofia and Yan, Tingting and Dalal, Ananya and
                 DiMartino, James M and Brown, Sam and Safo-Mensa, Nana and
                 Greenberg, Ethan and Cook, Michael and Finley-May, Samantha and
                 Flynn, Miriam A and Hopkins, Gary Patrick and Kovalyak, Julie
                 and Leonard, Meghan and Lohff, Alanna and Ordish, Christopher
                 and Scott, Ashley L and Takemura, Satoko and Walsh, Claire and
                 Walsh, John J and Berger, Daniel R and Pfister, Hanspeter and
                 Berg, Stuart and Knecht, Christopher and Meissner, Geoffrey W
                 and Korff, Wyatt and Ahrens, Misha B and Jain, Viren and
                 Lichtman, Jeff W and Engert, Florian",
  journal     = "bioRxivorg",
  institution = "bioRxiv",
  pages       = "2025.06.10.658982",
  month       =  jun,
  year        =  2025,
  language    = "en"
}

@ARTICLE{Motta2019-mq,
  title     = "Dense connectomic reconstruction in layer 4 of the somatosensory
               cortex",
  author    = "Motta, Alessandro and Berning, Manuel and Boergens, Kevin M and
               Staffler, Benedikt and Beining, Marcel and Loomba, Sahil and
               Hennig, Philipp and Wissler, Heiko and Helmstaedter, Moritz",
  journal   = "Science",
  publisher = "American Association for the Advancement of Science (AAAS)",
  volume    =  366,
  number    =  6469,
  pages     = "eaay3134",
  month     =  nov,
  year      =  2019,
  language  = "en"
}

@ARTICLE{Tang2024-ku,
  title     = "Optogenetic brain-computer interfaces",
  author    = "Tang, Feifang and Yan, Feiyang and Zhong, Yushan and Li, Jinqian
               and Gong, Hui and Li, Xiangning",
  journal   = "Bioengineering (Basel)",
  publisher = "MDPI AG",
  volume    =  11,
  number    =  8,
  pages     =  821,
  month     =  aug,
  year      =  2024,
  language  = "en"
}

@ARTICLE{Pedregosa2012-ec,
  title         = "Scikit-learn: Machine Learning in Python",
  author        = "Pedregosa, Fabian and Varoquaux, Gaël and Gramfort, Alexandre
                   and Michel, Vincent and Thirion, Bertrand and Grisel, Olivier
                   and Blondel, Mathieu and Müller, Andreas and Nothman, Joel
                   and Louppe, Gilles and Prettenhofer, Peter and Weiss, Ron and
                   Dubourg, Vincent and Vanderplas, Jake and Passos, Alexandre
                   and Cournapeau, David and Brucher, Matthieu and Perrot,
                   Matthieu and Duchesnay, Edouard",
  journal       = "arXiv [cs.LG]",
  month         =  jan,
  year          =  2012,
  archivePrefix = "arXiv",
  primaryClass  = "cs.LG"
}

@article{Adesnik2021-review,
title   = {Probing neural codes with two-photon holographic optogenetics},
author  = {Adesnik, Hillel and Abdeladim, Lamiae},
journal = {Nature Neuroscience},
volume  = {24},
number  = {10},
pages   = {1356--1366},
year    = {2021},
doi     = {10.1038/s41593-021-00902-9}
}

@article{Triplett2025-rapid,
title   = {Rapid learning of neural circuitry from holographic ensemble stimulation enabled by model-based compressed sensing},
author  = {Triplett, Marcus A. and Gajowa, Marta and Antin, Benjamin and Sadahiro, Masato and Adesnik, Hillel and Paninski, Liam},
journal = {Nature Neuroscience},
volume  = {28},
number  = {10},
pages   = {2154--2165},
year    = {2025},
doi     = {10.1038/s41593-025-02053-7}
}

@inproceedings{Wagenmaker2024-active,
title     = {Active learning of neural population dynamics using two-photon holographic optogenetics},
author    = {Wagenmaker, Andrew and Mi, Lu and Rozsa, Marton and Bull, Matthew S. and Svoboda, Karel and Daie, Kayvon and Golub, Matthew D. and Jamieson, Kevin},
booktitle = {Advances in Neural Information Processing Systems},
volume    = {37},
year      = {2024}
}

\appendix
\newpage
\section{Technical appendices and supplementary material}
\subsection{Derivation of Beta-Bernoulli connectivity mapping}

Recall that we assign the functional connection strength from a stimulated target $i$ to a putative responder neuron $j$ as a probability
\[
\theta_{ij} \in [0,1],
\]
interpreted as the probability that neuron $j$ shows a significant evoked response on a given trial, conditioned on photostimulation of $i$.

\paragraph{Bernoulli likelihood.}
On each trial $t = 1,\dots,T$, we summarize the response of neuron $j$ to
photostimulation of $i$ as a binary indicator
\[
s_t^{(ij)} =
\begin{cases}
1, & \text{if neuron $j$ is classified as ``responsive'' on trial $t$},\\
0, & \text{otherwise}.
\end{cases}
\]
For fixed $(i,j)$, we assume conditional independence across trials given
$\theta_{ij}$, so that
\[
s_t^{(ij)} \mid \theta_{ij} \sim \text{Bernoulli}(\theta_{ij}), \qquad t = 1,\dots,T.
\]
The likelihood of the entire set of observations
$\mathbf{s}^{(ij)} = \{ s_t^{(ij)} \}_{t=1}^T$ is then
\begin{align}
p\!\bigl(\mathbf{s}^{(ij)} \mid \theta_{ij}\bigr)
&= \prod_{t=1}^T p\!\bigl(s_t^{(ij)} \mid \theta_{ij}\bigr) \\
&= \prod_{t=1}^T
    \theta_{ij}^{\,s_t^{(ij)}} (1 - \theta_{ij})^{\,1 - s_t^{(ij)}} \\
&= \theta_{ij}^{\sum_{t=1}^T s_t^{(ij)}}
   \,(1 - \theta_{ij})^{T - \sum_{t=1}^T s_t^{(ij)}}.
\end{align}
For convenience, define the total number of ``successes'' (responsive trials)
\[
S_{ij} \;=\; \sum_{t=1}^T s_t^{(ij)}.
\]
Then the likelihood can be written compactly as
\begin{equation}
p\!\bigl(\mathbf{s}^{(ij)} \mid \theta_{ij}\bigr)
= \theta_{ij}^{S_{ij}} (1 - \theta_{ij})^{T - S_{ij}}.
\end{equation}

\paragraph{Beta prior.}
We place a Beta prior on the unknown response probability $\theta_{ij}$:
\[
\theta_{ij} \sim \text{Beta}(\alpha_0, \beta_0),
\]
with density
\begin{equation}
p(\theta_{ij})
= \frac{1}{B(\alpha_0,\beta_0)}
  \theta_{ij}^{\alpha_0 - 1} (1 - \theta_{ij})^{\beta_0 - 1},
\end{equation}
where $B(\alpha_0,\beta_0)$ is the Beta function. Intuitively, $\alpha_0$ and $\beta_0$ can be interpreted as prior pseudo-counts of successes and failures, respectively (e.g. with a weakly informative prior $\alpha_0=\beta_0=1$).

\paragraph{Posterior (conjugacy).}
By Bayes' rule, the posterior over $\theta_{ij}$ given the observed trials
$\mathbf{s}^{(ij)}$ is
\begin{align}
p\!\bigl(\theta_{ij} \mid \mathbf{s}^{(ij)}\bigr)
&\propto
p\!\bigl(\mathbf{s}^{(ij)} \mid \theta_{ij}\bigr)\, p(\theta_{ij}) \\
&\propto
\left[
  \theta_{ij}^{S_{ij}} (1 - \theta_{ij})^{T - S_{ij}}
\right]
\left[
  \theta_{ij}^{\alpha_0 - 1} (1 - \theta_{ij})^{\beta_0 - 1}
\right] \\
&\propto
\theta_{ij}^{(\alpha_0 - 1) + S_{ij}}
(1 - \theta_{ij})^{(\beta_0 - 1) + (T - S_{ij})}.
\end{align}
Recognizing the kernel of a Beta distribution, we obtain
\begin{equation}
\theta_{ij} \mid \mathbf{s}^{(ij)}
\;\sim\;
\text{Beta}\!\bigl(\alpha_0 + S_{ij},\; \beta_0 + T - S_{ij}\bigr).
\end{equation}

Thus the Beta prior is conjugate to the Bernoulli likelihood, and the posterior simply adds the observed counts of successes and failures to the prior pseudo-counts.

\paragraph{Posterior mean connectivity estimate.}
A convenient point estimate of the functional connection strength from $i$ to $j$ is the posterior mean
\begin{align}
\hat{\theta}_{ij}
&:= \mathbb{E}\bigl[\theta_{ij} \mid \mathbf{s}^{(ij)}\bigr] \\
&= \frac{\alpha_0 + S_{ij}}{\alpha_0 + \beta_0 + T}.
\end{align}
For the common choice of a uniform prior $\alpha_0 = \beta_0 = 1$, this reduces to
\begin{equation}
\hat{\theta}_{ij}
= \frac{1 + S_{ij}}{2 + T},
\end{equation}
We use $\hat{\theta}_{ij}$ as the final edge weight in our probabilistic functional connectome, representing the estimated probability that neuron $j$ responds to photostimulation of neuron or ensemble $i$.

\subsection{Collection and generation of data}\label{ssec:datagen}
\paragraph{Synthetic standalone stimulation trials}
To generate heterogenous synthetic connectomes, we sample stimulation rows from a Bernoulli Mixture Model (BMM), where each target $i$ belongs to one of several latent response classes (low, high, ambiguous). Row probabilities are drawn conditional on class membership. Given unknown causal probability $\theta_{ij}$ and responder $j$, each stimulation row $i$ is first assigned a latent class:

\[
z_i \sim \mathrm{Categorical}(\pi),
\]
corresponding to low-probability, high-probability, or ambiguous response regimes. Conditional on class membership, edge probabilities are sampled from component-specific distributions,
\[
\theta_{ij}\mid z_i=k \sim p_k,
\]
after which trial outcomes are generated as
\[
r_{ij}^{(\ell)} \sim \mathrm{Bernoulli}(\theta_{ij}).
\]

where $r_{ij}^{(\ell)}=1$ indicates an evoked response and $0$ otherwise. Across $N_i$ repeated stimulations of target $i$, the empirical estimate is
\[
\hat \theta_{ij}
=
\frac{1}{N_i}\sum_{\ell=1}^{N_i} r_{ij}^{(\ell)}.
\]

For large $N_i$, this constitutes a comprehensive connectome.

\paragraph{Experimental standalone stimulation trials}\label{sssec:exp_criteria}
For experimentally-derived connectomes, per each stimulation target $i$ and responder $j$, we measure raw changes in calcium fluorescence in traced cells following holographic photostimulation. Using a baseline defined as the mean fluorescence, $f_0$, we compute $\Delta f/f$ or $(f-f_0)/f_0$, then z-score that data to get $z_{ij}^{(\ell)}$ for trial $\ell$. Using a trial-centered baseline (14 frames) and post-stimulation windows (8 frames), we threshold each response as:
\[
r_{ij}^{(\ell)}
=
\mathbb{I}[z_{ij}^{(\ell)}>\tau], \text{ where } \tau = \mu_{\textrm{baseline}}+\Gamma(\sigma_{\textrm{baseline}})
\]

for pre-defined multiplier threshold $\Gamma$. For all trials, we consider $\Gamma = 1.3$ (determined empirically).

For standalone data collection, experiments were performed on 6-9 days post-fertilization (dpf) zebrafish larvae expressing the double-transgenic red-shifted excitatory opsin rsChRmine \cite{Kishi2022-mi} under the HuC/elavl3 promoter \cite{Park2000-bm, Khuansuwan2019-qb} as well as the green calcium indicator GCaMP6s\cite{Chen2013-nb}. All fish were head fixed in 1.5\% low melting point agarose. Volumetric two-photon calcium imaging (Bruker Ultima, 920 nm, $<20$ mW) was performed spanning the visually responsive pretectum and adjacent hindbrain areas, over $512\times512$ pixels ($\sim140\,\mu$m field), 3 planes (8–10 $\mu$m spacing) at 3.4 Hz. Prior to optogenetic photostimulation, to characterize neural spontaneous and visual response properties, all zebrafish were exposed to repetitions of moving grating stimuli which evoke optomotor response behaviors \cite{Fouke2025-xy}. Visual stimuli were presented from below in 4 randomly interleaved  repetitions with a 23 second baseline and 7 seconds of motion. 10 optic flow stimuli were shown: binocular converging, diverging, left, right, forward, backward, and monocular medial-left, medial-right, lateral-right, and lateral-left \cite{Naumann2016-mx}. 

Neural source regions of interests (ROI) were extracted from fluorescence images using Suite2P \cite{Pachitariu2016-vp}, and stimulation targets (coordinates of single cells and ensembles of size 1,3,7,10; \(>10\,\mu\)m spacing) were selected based on visual response properties. To perform photostimulation, neurons were targeted with a 1035 nm femtosecond laser (SLM holography; \(\sim\)30 mW per target, 0.2 second spirals), in randomized events (up to 10 repeats, 20–30 second spacing). Motion correction (NoRMCorre \cite{Pnevmatikakis2017-qu}, CaImAn \cite{Giovannucci2019-ix}) and source extraction were performed post hoc. Prior features were extracted from spontaneous and visually evoked neural activity using population correlation statistics (e.g., mean, variance, quantiles). Additionally, we constructed synthetic and hybrid combinatorial datasets from experimentally measured ensemble responses (Sec. \ref{ssec:datagen}).

\paragraph{Experimental combinatorial stimulation trials}\label{sssec:exp_comb_stim}
Combinatorial experiments were performed in three zebrafish using five candidate target cells per fish. We stimulated all non-empty target combinations except pairs, which could not be reliably generated with our holographic photostimulation configuration \cite{Pegard2017-tb}, yielding 21 combinations per fish. For each fish, 12 trials per combination were used to construct the trial-exhaustive reference connectome, and each algorithm was evaluated using a specified proportion of these trials.

Experimental parameters were otherwise identical to the standalone experiments, except that no visual stimuli were presented. Each combination was stimulated once before repeating the full set, with alignment performed between trials. Spontaneous activity collected before photostimulation was used as prior information for OPh-AL and OPh-CS, and cells were aligned post hoc between spontaneous and stimulation datasets. More details will be made available in an associated code repository. 

\subsection{Strategies for multi-armed bandits}
For standalone action selection of neural targets, we use OPhELIA to augment Bayesian Active Learning (BAL). We also consider a selection of well-known and widely used solutions to the multi-armed bandit problem, which we describe below as they are framed in the context of optimal photostimulation selection of ensembles.

\subsubsection{Bayesian Active Learning (BAL)}\label{sssec:bal}
For the first model that we augment with OPhELIA, we employ a first order approximation of Bayesian Active Learning by Disagreement (BALD) \cite{Houlsby2011-my}, which is a prominent active learning strategy which is well-suited for selecting informative training set samples for supervised learning problems \cite{Werner2025-kd}. In our case, the goal is, equivalently, to select the next stimulation target $i$.

In a BALD formulation for our problem, the optimal query would maximize the expected reduction in posterior entropy. However, notice that $
\mathbb{E}\left[ \mathcal{H}[p(r \mid i, \theta)] \right]$ is intractable, since response vector $r \in \{0,1\}^{\mathcal{N}_r}$ entails a summation over $2^{\mathcal{N}_r}$ possible outcomes. Instead of the BALD objective term, we thus consider the posterior variance, and select actions which minimize this term. With OPhELIA, we refer to our BAL approach as OPh-AL.

\subsubsection{Thompson sampling}\label{sssec:ts}
As a stochastic alternative to OPhELIA, we employ Thompson sampling. For each edge, we sample:
\begin{equation}
\tilde{\theta}_{ij} \sim \mathrm{Beta}(\alpha_{ij}, \beta_{ij}),
\end{equation}
and define the sampled row utility:
\begin{equation}
\tilde{U}(i) = \sum_{j=1}^{\mathcal N_r} \tilde{\theta}_{ij}.
\end{equation}
We then select:
\begin{equation}
i_t = \arg\max_i \tilde{U}(i).
\end{equation}

This procedure samples from the posterior over connectomes and greedily selects the most promising stimulation under that sample, naturally balancing exploration and exploitation.

\subsubsection{Upper confidence bound (UCB)}\label{sssec:ucb}
We also consider an upper confidence bound strategy. For each edge, we define:
\begin{equation}
\mathrm{UCB}_{ij} = \hat{\theta}_{ij} + c \cdot \sqrt{\mathrm{Var}[\theta_{ij}]},
\end{equation}
where $c > 0$ controls exploration. For all model comparisons, we use a fixed $c=2.0$.

The row-level utility is:
\begin{equation}
U(i) = \sum_{j=1}^{\mathcal N_r} \mathrm{UCB}_{ij},
\end{equation}
and we select:
\begin{equation}
i_t = \arg\max_i U(i).
\end{equation}

This encourages sampling of stimulations with high estimated connectivity and/or high uncertainty.

\subsubsection{Round-robin stimulation}\label{sssec:rr}
As a baseline, we consider a uniform sampling strategy that cycles through all stimulation targets:
\begin{equation}
i_t = t \bmod \mathcal N_s.
\end{equation}

This ensures equal sampling across all presynaptic cells but does not adapt to uncertainty or information gain.

\subsection{Strategies for combinatorial optimization}
\subsubsection{Compressed Sensing (CS)}\label{sssec:cs}
Of the two approaches which OPhELIA serves to augment, we use Compressed Sensing (CS) for combinatorial selection. We consider a variant of CS which (1) is a generative model, rather than a reconstructive one; and (2) frames action selection using our previously derived ambiguity score (Eq. \ref{eqn:ambig}). 

CS assumes that multi-cell target responses can be reconstructed from a sparse set of latent single-cell contributions. Let $\mathcal{V} = \{1,\dots,\mathcal{N}_s\}$ denote stimulation targets and $\mathcal{C}$ the set of candidate subsets. We define a binary design matrix
$ \mathbf{C} \in \{0,1\}^{|\mathcal{C}| \times \mathcal{N}_s}$, 
where each row $\mathbf{c}_k$ encodes a subset $C_k \in \mathcal{C}$. For each responder $j$, we train a simple linear model with an $\ell_1$-norm penalty:
\begin{equation}
\hat{P}_{kj}
=
\mathbf{c}_k^\top w_j + b_j,
\rightarrow
\hat{w}_j
=
\arg\min_{w_j}
\left\|
\mathbf{y}_{j,\mathrm{obs}} - \mathbf{C}_{\mathrm{obs}} w_j
\right\|_2^2
+
\lambda \|w_j\|_1.
\end{equation}

Here, $\hat{P}_{kj}$ is the predicted response probability of responder $j$ to subset $C_k$, $\mathbf{c}_k \in \{0,1\}^{\mathcal{N}_s}$ is the binary design vector encoding subset $C_k$; $w_j \in \mathbb{R}^{\mathcal{N}_s}$ are the latent stimulation weights for responder $j$; $b_j$ is a bias term; $\mathbf{C}_{\mathrm{obs}}$ is the design matrix restricted to observed stimulation subsets; $\lambda = 0.01$ is a regularization parameter; and $\mathbf{y}_{j,\mathrm{obs}}$ is the vector of observed response probabilities of responder $j$.

Unobserved subset responses are generated using a $\mathrm{clip}$ nonlinearity for Beta parameter updates.
\begin{equation}
\hat{P}_{kj}
=
\mathrm{clip}\!\left(
\mathbf{c}_k^\top \hat{w}_j + \hat{b}_j,
\epsilon, 1-\epsilon
\right)
\rightarrow
\hat{\alpha}_{kj}
=
\alpha_0 + n_{\mathrm{eff}} \hat{P}_{kj},
\qquad
\hat{\beta}_{kj}
=
\beta_0 + n_{\mathrm{eff}} (1 - \hat{P}_{kj}),
\end{equation}

where $n_{\textrm{eff}}=8$ is the effective sample size assigned to reconstructed observations. This enables propagation of sparse reconstruction into the Bayesian posterior over connectivity. Feeding this into our Bayesian sequential inference loop, we next use the predicted responses for unobserved subsets to guide exploration. When augmented with OPhELIA, we refer to this model as OPh-CS.

\subsubsection{Noisy group testing (NGT)}\label{sssec:NGT}
We adapt the NGT framework of Draelos et al. \cite{Draelos2020-sd,Draelos2020-zj} to our probabilistic connectivity objective and experimental design, rather than reproducing their original method exactly. We model pooled stimulation via a noisy-OR likelihood over single-edge probabilities $\{\theta_{ij}\}$. For a stimulation set $S_t \in \mathcal{C}$ with indicator vector $\mathbf{s}_t \in \{0,1\}^{\mathcal{N}_s}$, responses follow
\begin{equation}
r_{tj} \sim \mathrm{Bernoulli}\!\left(
1 - \prod_{i=1}^{\mathcal{N}_s} (1 - \theta_{ij})^{(\mathbf{s}_t)_i}
\right).
\end{equation}
Given observed trials $\{(S_\tau, r_{\tau j})\}_{\tau=1}^t$, we estimate $\theta_{ij}$ by maximizing
\begin{equation}
\hat{\theta}_{\cdot j}^{(t)}
=
\arg\max_{\theta_{\cdot j} \in [0,1]^{\mathcal{N}_s}}
\sum_{\tau=1}^t
\left[
r_{\tau j} \log p_{\tau j}
+
(1-r_{\tau j}) \log(1-p_{\tau j})
\right],
\end{equation}
where
\begin{equation}
p_{\tau j}
=
1 - \prod_{i=1}^{\mathcal{N}_s} (1 - \theta_{ij})^{(\mathbf{s}_\tau)_i}.
\end{equation}
Predicted probabilities $\hat{p}_{tj}$ are incorporated into the Beta posterior via pseudo-counts:
\begin{equation}
\alpha_{tj}^{(t)} = \alpha_0 + n_{\mathrm{eff}} \hat{p}_{tj},
\quad
\beta_{tj}^{(t)} = \beta_0 + n_{\mathrm{eff}} (1 - \hat{p}_{tj}),
\end{equation}
and selection uses $U(S_t)=\sum_j (1-2|\hat{p}_{tj}-0.5|)\mathrm{Var}[\theta_{tj}^{(t)}]$.

\vspace{0.2cm}

\subsubsection{Low-rank matrix recovery (LRM)}\label{sssec:LRM}

We assume $\Theta \in [0,1]^{\mathcal{N}_s \times \mathcal{N}_r}$ is low-rank, parameterized as
\begin{equation}
\theta_{ij} = \sigma(u_i^\top v_j),
\quad
U \in \mathbb{R}^{\mathcal{N}_s \times d},\;
V \in \mathbb{R}^{\mathcal{N}_r \times d},
\end{equation}
with $d \ll \min(\mathcal{N}_s,\mathcal{N}_r)$. Here we used $d=5$. Observations follow
\begin{equation}
r_{ij}^{(\tau)} \sim \mathrm{Bernoulli}(\theta_{ij}),
\end{equation}
and parameters are estimated from $\mathcal{D}^{(t)}$ via
\begin{equation}
\min_{U,V}
\sum_{(i,j,\tau)\in\mathcal{D}^{(t)}}
\left[
- r_{ij}^{(\tau)} \log \sigma(u_i^\top v_j)
- (1-r_{ij}^{(\tau)}) \log (1-\sigma(u_i^\top v_j))
\right]
+
\lambda(\|U\|_F^2+\|V\|_F^2).
\end{equation}
Predictions $\hat{\theta}_{ij}^{(t)}=\sigma(\hat{u}_i^\top \hat{v}_j)$ are mapped to Beta updates:
\begin{equation}
\alpha_{ij}^{(t)}=\alpha_0+n_{\mathrm{eff}}\hat{\theta}_{ij}^{(t)},
\quad
\beta_{ij}^{(t)}=\beta_0+n_{\mathrm{eff}}(1-\hat{\theta}_{ij}^{(t)}),
\end{equation}
and selection uses $U(i)=\sum_j (1-2|\hat{\theta}_{ij}^{(t)}-0.5|)\mathrm{Var}[\theta_{ij}^{(t)}]$.

\subsection{Training a random forest for prior inference}\label{app:rf_prior}

To predict target-level ambiguity from pre-perturbation activity
features, we used a \texttt{RandomForestRegressor} (Scikit-learn) \cite{Pedregosa2012-ec} with \texttt{n\_estimators}=300, \texttt{max\_depth}=4, \texttt{min\_samples\_leaf}=2. Input features were derived from spontaneous and visually evoked activity statistics prior to photostimulation experiments. The model was trained to predict mean responder entropy
\(\bar{\mathcal H}_i\) for each stimulation target \(i\) or $C_k$ for combinatorial selection. We weighted the learned prior contribution using parameter $\lambda=0.35$. Model performance was evaluated using leave-one-out cross-validation (LOOCV), and performed separately for each subject. 

\subsection{LLM usage}
We used OpenAI GPT-5.6 for editing the manuscript writing and math checking, and Claude Opus 4.8 for figure brainstorming (e.g. parts of Figures 0 and 1) and generating code for plotting.

\subsection{Compute resources used}\label{sssec:compres}
For analysis of all experiments, we used a Linux-based PC with an AMD Ryzen 7 7800X3D 8-Core Processor and two NVIDIA RTX 3080 GPUs, each with 24GB of GDDR6X VRAM.

\end{document}